\documentclass[12pt]{iopart}
\usepackage[dvips]{graphicx}
\usepackage{amssymb}
\usepackage{amsfonts}

\begin{document}

\title{Exotic fluids and crystals of soft polymeric colloids}

\author{Christos N Likos{\dag}, Norman Hoffmann{\dag}, 
Hartmut L{\"o}wen{\dag} and
Ard A Louis\ddag}
\address{\dag Institut f{\"u}r Theoretische Physik II,
Heinrich-Heine-Universit\"at D{\"u}sseldorf,
Universit\"atsstra{\ss}e 1, D-40225 D\"usseldorf, Germany\\
\ddag Department of Chemistry, University of Cambridge, Lensfield Road,
Cambridge CB2 1EW, United Kingdom}
\date{\today}

\begin{abstract}
We discuss recent developments and present new findings
in the colloidal description of soft
polymeric macromolecular aggregates.  For various macromolecular
architectures, such as linear chains, star polymers, dendrimers and
polyelectrolyte stars, the effective interactions between suitably
chosen coordinates are shown to be ultrasoft, i.e., they either remain
finite or diverge very slowly at zero separation. As a consequence,
the fluid phases have unusual characteristics, including anomalous
pair correlations and mean-field like thermodynamic behaviour. The
solid phases can exhibit exotic, strongly anisotropic as well as open
crystal structures. For example, the diamond and the A15-phase are
shown to be stable at sufficiently high concentrations.  Reentrant
melting and clustering transitions are additional features displayed
by such systems, resulting in phase diagrams with a very rich
topology. We emphasise that many of these effects are fundamentally
different from the usual archetypal hard sphere paradigm.  Instead,
we propose that these fluids fall into the class of mean-field fluids.
\end{abstract}



\section{Introduction}
\label{intro.sec}

One major advantage of soft matter systems in comparison to their atomic
counterparts is the ability one has to engineer the constituent particles
at the molecular level. In this way, an enormous variety in architectures
can be achieved, leading to a corresponding richness in the structural and
phase behaviour of such systems. Polymers play a prominent example 
within this class of materials. They come in a variety of forms, such as, 
e.g., as linear chains, branched, star-shaped, dendritic, copolymer, 
as well as functions, e.g., steric stabilisers, additives, depletants etc.
Additional flexibility arises from the
possibility of influencing the structural and phase behaviour of
polymer solutions by changing the solvent quality.

From the theoretical point of view, the task of bridging the gap between
the microscopic and macroscopic length scales of polymeric systems is
a formidable one. Indeed, the constituent 
macromolecules may contain thousands or even millions of 
atoms, interconnected to one another in complicated ways. The program of
starting with the individual interactions between monomers and 
proceeding to the calculation of the free energy of the system 
rapidly becomes intractable.
However, considerable progress can be made if one
invokes a ``coarse-graining'' procedure, a point of view that has been
proved very fruitful in many areas of research in condensed matter
physics.  Here we follow a two-step procedure. First, instead of
attempting to carry out the statistical trace over all the individual
monomers in one step, a certain generalised coordinate of the
macromolecule as an effective point particle is invoked. Examples
include the centre of mass of the polymer or some suitably selected
monomer, as we explain below. All monomers belonging to the
macromolecules are then traced out for a given, fixed configuration of
the effective coordinates, which defines an {\it effective
interaction} between these coordinates \cite{pusey:leshouches:89,
likos:physrep:01}.  Once this is achieved, the second step consists of
viewing the the macroscopic system as collection of point particles
interacting by means of the effective interaction. Now all known tools
from the theory of atomic and molecular fluids can be employed to
derive structural and thermodynamic quantities for the system under
consideration.

In the recent years, the program sketched above has been carried out
with success for various polymeric systems. It has been found that
the effective interactions obtained belong to a new class of ultrasoft
potentials which have very unusual properties when compared with the
hard-sphere (HS) system, the prototype of atomic liquids. In this work,
we first present a concise review of these novel properties for the 
fluid phases and then some new results regarding the rich variety
of crystal structures that can be stabilised by ultrasoft potentials.
The rest  of the paper is organised as follows. In section
\ref{effective.section} we give the general definition of the effective
interaction and present specific examples for a number
of systems that have been worked out recently.
In section \ref{mff.section} we show that the systems described by
ultrasoft interactions are well described by a simple mean-field
picture in the fluid phase  for a wide range of thermodynamic
conditions.
To further delineate these systems we define two
categories of mean-field fluids, the {\it strong} and the {\it weak}
ones. The former are characterised by a direct correlation
function $c(r)$ that satisfies to excellent accuracy the condition
$c(r) = -\beta v(r)$ in broad regions of the thermodynamics space,
where $v(r)$ is the interparticle pair potential. The latter
only  satisfy an approximate mean-field picture for their thermodynamic
properties, and not for their structure.
In section \ref{cluster.section}
we demonstrate that strong mean-field fluids can be further divided into
two categories: those displaying reentrant melting and those displaying
a cascade of clustering transitions, 
the criterion being set by the positivity of
the Fourier transform of the effective interaction. 
In section \ref{exotic.section}
we discuss the richness of the phase diagrams of weak mean-field fluids,
taking the case of star polymers as a concrete example. Reentrant melting
as well as a wealth of structural phase transitions and exotic crystal phases
are all shown to be stemming from the ultrasoftness of the effective interactions.
Finally, in section \ref{summary.section} we summarise and conclude.

\section{Effective interactions between polymeric macromolecules}
\label{effective.section}

The effective interaction between flexible, fluctuating aggregates
can be given a precise statistical-mechanical definition.
Let us consider a solution containing $M$ polymeric macromolecules,
each one of them composed of $N$ monomers. The total number of
particles in the system is ${\mathcal M} = M \times N$. 
One starts from the full Hamiltonian ${\mathcal H}$ of the problem, assumed 
to be known. Then, out of the ${\mathcal M}$ particles in the problem 
(in our case all monomers), 
one selects the $M$ ones that are to be considered as ``effective particles'' 
and holds them fixed in some prescribed configuration 
$\{{\bf R}_1, {\bf R}_2, \ldots, {\bf R}_M\}$, where ${\bf R}_i$ 
is the position of the $i$-th effective particle. 
Afterwards, the ${\mathcal M} - M$ remaining particles 
are canonically traced out and the result of this integration 
is a constrained partition function $Q({\bf R}_1, {\bf R}_2, \ldots, {\bf R}_M)$. 
The effective Hamiltonian ${\mathcal H}_{\rm eff}$ is defined as:
\begin{equation}
\exp\left(-\beta{\mathcal H}_{\rm eff}\right) =
Q({\bf R}_1, {\bf R}_2, \ldots, {\bf R}_M),
\label{heff.eq}
\end{equation}
where $\beta = (k_BT)^{-1}$, with the absolute temperature $T$ and
Boltzmann's constant $k_B$. It can be shown \cite{likos:physrep:01}
that such an effective Hamiltonian has two useful
properties:
it preserves the overall thermodynamics of the system and it
guarantees that the correlation functions of any order between
any of the $M$ remaining particles remain invariant, regardless
of whether the expectation values are calculated with the original
Hamiltonian ${\mathcal H}$ or with
the effective one ${\mathcal H}_{\rm eff}$.
Though the procedure of tracing out
the ${\mathcal M} - M$ degrees of freedom necessarily                           
generates interactions of all orders between the 
$M$ particles \cite{dijkstra:pre:99, dijkstra:jpcm:99}
in many cases it is sufficient to truncate those at the pair level,
introducing thereby the {\it pair-potential approximation}. The great advantage
of employing this point of view is that, in comparison with the 
original problem, the numbers of particles,
${\mathcal M}$, has been reduced by a factor $N$. In addition, whereas
in the original problem the pair interactions between the monomers
are quite complicated, due to the need of taking into account the
connectivity and architecture of the molecule, in the effective description
the pair potential is spherically symmetric, depending only on the 
magnitude of the separation vector between the two effective coordinates.
A new picture of the polymers emerges thereby in which the
latter can be seen as ultrasoft colloids having a dimension of the
order of their radius of gyration $R_g$. This sets at the same time the 
characteristic length scale of the effective interaction between them.
When the polymers are neutral, $R_g$ is the only length scale appearing
in this colloidal description, whereas if they carry charge, additional
scales set, e.g., by the concentration of the solution, the counterions
and/or the salt ions also come into play.
We examine some characteristic cases below.

{\it Polymer chains.} Two possible choices for the effective coordinates have
been investigated thus far. One possibility is to consider the effective
interaction $v_{\rm com}(r)$ between the {\it centres of mass} of the linear
chains, when these are kept fixed at a distance $r$ from one another. 
This was first done in the
pioneering work of Flory and Krigbaum \cite{flory:krigbaum:50}, who
found a Gaussian interaction between the centers of mass.  Although
the functional form of the Flory-Krigbaum potential is correct, the
dependence of the prefactor on the degree of polymerisation $N$ is
not: whereas the Flory-Krigbaum mean-field approach predicts a
$N^{1/5}$-dependence the prefactor turns out to be $N$-independent for
sufficiently large $N$, as can be easily seen from examining the
polymer second virial coefficient.  Standard scaling arguments show
that the radius of gyration $R_g$ is the only relevant length scale for
the dilute and semi-dilute regimes of polymers in a good
solvent \cite{deGennes:79}.  This immediately implies that the
second-virial coefficient should scale as $B_2 \sim R_g^3$, which, in
turn, implies that effective interaction must have an amplitude that
is independent of $R_g$ \cite{ard:all:02}, at least in the scaling
limit.

A number of 
simulational \cite{grosberg:82, schaefer:baumgaertner:86, dautenhahn:hall:94}
as well as theoretical approaches \cite{krueger:etal:89} involving
two self-avoiding chains, reached the conclusion that the aforementioned
interaction has a Gaussian form. The lack of divergence of this effective
interaction at zero separation should not be surprising. Indeed, the centres
of mass of two polymer chains can coincide without any of the monomers
violating the excluded volume conditions. In addition, it can be seen
that the effective ``particles'' one chooses for the coarse-grained
description of the system do not need to be real particles of the physical
system. 

Recently, Louis {\it et al.}\ \cite{ard:peter:00, ard:mft:00, bolhuis:jcp:00}
have independently carried out
state-of-the art simulations involving not just two but $N_c$ chains
and varying the number of chains to cover a very broad range of
concentrations, ranging from dilute solutions up to nine times
the overlap concentration. They confirmed that the
effective potential has a Gaussian-like form which at zero density can
be well approximated by
\begin{equation}
v_{\rm com}(r) = \varepsilon\exp[-(r/\sigma)^2], 
\label{vcom.eq}
\end{equation}
where $\varepsilon = 1.87\,k_BT$
and $\sigma = 1.08\,R_g$. For
higher densities a superposition of three Gaussians provides a very
accurate fit \cite{bolhuis:macro:02}, but the basic shape does not
deviate much from the low-density Gaussian form.  The rather weak density
dependence can be shown to arise from the density independent
many-body forces \cite{bolhuis:pre:01}.
Moreover, the same authors have shown
that employing this effective interaction leads to a very accurate
description of the thermodynamics (equation of state) of polymer
solutions for a wide range of concentrations, thus confirming the
validity of the idea that polymer chains can be viewed as
soft colloids \cite{bolhuis:jcp:00}. 

An alternative is to consider the end-monomers or the central monomers
of the chains as effective coordinates \cite{ard:all:02}. 
General scaling arguments
establish that in this case the effective interaction diverges 
logarithmically with the monomer-monomer 
separation $r$ \cite{witten:pincus:86:2}. 
When the central monomers are chosen, linear chains are equivalent to 
star polymers with $f = 2$ arms. Motivated by this analogy, Jusufi
{\it et al.}\ \cite{jusufi:etal:jpcm:01}
derived the effective interaction $v_{\rm cm}(r)$ by
combining monomer-resolved, off-lattice simulations with theoretical
arguments. The sought-for interaction features in this case a logarithmic
divergence for small separations, in full agreement with the scaling arguments,
and crosses over to a Gaussian decay for larger ones:
\begin{eqnarray}
v_{\rm cm}(r)  = \frac{5}{18}k_BTf^{3/2}
 \cases{
   -\ln\left(\frac{r}{\sigma_{\rm s}}\right)+\frac{1}{2\tau^{2}\sigma_{\rm s}^{2}} &
   {\rm for $r \leq \sigma_{\rm s}$};
 \\
   {\frac{1}{2\tau^{2}\sigma_{\rm
 s}^{2}}\exp\left[-\tau^{2}(r^{2}-\sigma_{\rm s}^{2})\right] } &
   {\rm for $r > \sigma_{\rm s}$},
 }
\label{pot_ss2}
\end{eqnarray}
where $\sigma_{\rm s} \cong 0.66 R_g$ and
$\tau(f)$ is a free parameter of the order of  $1/R_g$ and is
obtained by fitting to computer simulation results.
For $f = 2$ the value $\tau\sigma_{\rm s} = 1.03$ has been 
obtained \cite{jusufi:etal:jpcm:01},
which, together with the potential in Eq.\ (\ref{pot_ss2}) above
yields for the
second virial coefficient of polymer solutions                                            the value $B_2/R_{\rm g}^3 = 5.59$, in agreement with the estimate
$5.5 < B_2/R_{\rm g}^3 < 5.9$ from renormalisation group
and simulations \cite{bolhuis:jcp:00}. 

{\it Dendrimers.} By employing a simple, mean-field theory based on
the measured monomer density profiles of fourth-generation dendrimers,
a Gaussian function of the form (\ref{vcom.eq}) has
been shown to accurately describe the effective interaction
between the centres of mass of these dendrimers \cite{likos:01, likos:02}.
The prefactor $\varepsilon$ has in this case a higher value
then for linear polymers, $\varepsilon \cong 10\,k_BT$.
Small-angle neutron scattering (SANS) profiles from concentrated
dendrimer solutions are reproduced very well theoretically,
at least below the overlap concentration $c_{*}$.

{\it Star polymers.} By chemically anchoring $f$ linear chains on a common
core, star polymers with functionality $f$ are constructed. In the theoretical
analysis of the conformations and the effective interactions of stars,
the finite size of the core particle is ignored, an excellent approximation
when the chains are long. The natural choice for the effective coordinates
is now the position of the central particle, i.e., of the star centre.
For small functionalities, $f \lesssim 10$, 
Jusufi {\it et al.}\ \cite{jusufi:etal:jpcm:01} have
shown that a logarithmic-Gauss potential of Eq.\ (\ref{pot_ss2})
accurately describes the effective interaction. The decay parameter
$\tau$ of the Gaussian is $f$-dependent, for details see 
Ref.\ \cite{jusufi:etal:jpcm:01}. For larger functionalities, $f \gtrsim 10$,
the Daoud-Cotton \cite{Daoud:Cotton:82:1} 
blob picture of the stars is valid and the
star-star interaction potential $v_{\rm ss}(r)$
reads as \cite{likos:etal:prl:98, jusufi:macromolecules:99}:
\begin{eqnarray}
v_{\rm ss}(r)  = \frac{5}{18}k_BTf^{3/2}
\cases{
   -\ln\left(\frac{r}{\sigma_{\rm s}}\right) + \frac{1}{1+\sqrt{f}/2}
   & {\rm for $r \leq \sigma_{\rm s}$};
   \\
   {\frac{\sigma_{\rm s}/r}{1+\sqrt{f}/2}
    \exp\left[-\frac{\sqrt{f}}{2\sigma_{\rm s}}(r-\sigma_{\rm s})\right]}
   & {\rm for $r > \sigma_{\rm s}$},
}
\label{pot_ss}
\end{eqnarray} 
with the ``corona diameter'' $\sigma_{\rm s} \cong 0.66\,R_g$. Both star-star
potentials, the one valid for $f \lesssim 10$, Eq.\ (\ref{pot_ss2}) and
the one valid for $f \gtrsim 10$, Eq.\ (\ref{pot_ss}), show an ultrasoft
logarithmic divergence as $r \to 0$. The strength of the divergence is
controlled by the functionality $f$, so that at the formal limit $f \to \infty$
the interaction (\ref{pot_ss}) tends to the HS-potential. 

{\it Polyelectrolyte stars.} If the polymer chains of a star polymer contain
ionisable groups, the latter dissociate upon solution in a polar (aqueous)
solvent, leaving behind charged monomers and resulting in a solution
consisting of charged star polymers and counterions. 
The resulting macromolecules
are called polyelectrolyte (PE) stars. In PE stars the chains are stretched
due to the Coulomb repulsion of the charged monomers. The degree of stretching
and condensation of counterions on the rods depends on the amount of charge
and on the Bjerrum length. For moderate to high charging fractions,
the effective interaction between the centres of the PE-stars have been
analysed recently by means of computer simulations and 
theory \cite{arben:prl:02, arben:jcp:02}. This interaction, 
$v_{\rm pes}(r)$, is dominated by the entropic contribution of the
counterions that remain trapped within the star corona. For a broad
range of functionalities and charge fractions, it can be accurately
described by the fit:
\begin{eqnarray}
\fl
\frac{v_{\rm pes}(r)}{k_BT} = \frac{\tilde{C}fN_{c}}{1-\zeta}\cases{
      \left[1-\left(\frac{r}{\sigma}\right)^{1-\zeta}\right] +
      \frac{2}{5}\left[\left(\frac{r}{\sigma}\right)^{2-\zeta} -1\right]
    + \frac{3(1-\zeta)}{5(1+\kappa \sigma)} &  {\rm {for $r\leq \sigma$}};
\\
      \frac{3(1-\zeta)}{5(1+\kappa \sigma)} \left(\frac{\sigma}{r}\right)
       \exp[-\kappa(r-\sigma)] &
{\rm {for $r\geq \sigma$}},
}
\label{Vpes.eq}
\end{eqnarray}       
where $N_c$ is the number of counterions, $\sigma$ the corona diameter of
the PE-star, and $\kappa$ the inverse Debye screening length due to the
free counterions. Finally, $\tilde C$ and $\zeta$ are fit parameters,
where $0 < \zeta < 1$. The last condition ensures that the potential
of Eq.\ (\ref{Vpes.eq}) above tends to a finite value as $r \to 0$. 
Hence, once more we are dealing with an ultrasoft interaction that
varies slowly as the particle centres approach one another.

Ultrasoft interactions therefore describe quite a number of different
systems. Their  common characteristic is that the constituent
particles are polymers of various architectures that dominate the
spatial extent of the aggregates. In other words, one expects similar
ultrasoft interactions to show up also when one deals with core-shell
particles, consisting of a solid core and a polymeric shell, whenever
the thickness of the latter greatly exceeds the radius of the former.
In addition, the ultrasoft interactions can be tuned
by controlling the number of arms, the
charge, the length of the chains, the generation number (in the case
of dendrimers) etc. Hence, it is useful to explore the general 
characteristics of this family of potentials and the ramifications
they have on the structural and thermodynamic properties of the fluid
and crystal phases of such systems.

\section{Mean-field fluids}
\label{mff.section}

Motivated by the fact that effective interactions between polymeric
colloids can be {\it bounded} (i.e., finite at all separations $r$),
we examine here in general the properties of systems characterised
by pair potentials of the form
\begin{equation}
v(r) = \varepsilon\phi(r/\sigma),
\label{bounded.eq}
\end{equation}
with $\phi(x) < \infty$ for all $x$. In Eq.\ (\ref{bounded.eq}) above,
$\varepsilon$ is an energy scale and $\sigma$ a length scale.
Moreover, $v(r)$ is non-attractive, i.e., 
${\rm d}\phi(x)/{\rm d}x \leq 0$ everywhere. 
We introduce dimensionless measures of temperature and density
as
\begin{eqnarray}
t & = & {{k_BT}\over{\varepsilon}} = \left(\beta\varepsilon\right)^{-1};
\\
\eta & = & {{\pi}\over{6}}\rho\sigma^3  =  {{\pi}\over{6}}\bar\rho,
\label{params}
\end{eqnarray}
where $k_B$ is Boltzmann's constant and $\rho = N/V$ is the density of
a system of $N$ particles in the volume $V$. We will refer to $\eta$
as the `packing fraction' of the system.

The key idea for examining
the high temperature and/or high density limit of such model systems
in three and higher dimensions is the following.
We consider in general
a spatially modulated density profile $\rho({\bf r})$ which does
not vary too rapidly on the scale $\sigma$ set by the interaction.
At high densities, $\rho\sigma^3 \gg 1$, the average interparticle
distance $a \equiv \rho^{-1/3}$ becomes vanishingly small,
and it holds $a \ll \sigma$, i.e., the potential is extremely
long-range. Every particle is simultaneously interacting with
an enormous number of neighboring molecules and in the absence of
short-range excluded volume interactions the excess free energy
of the system \cite{evans:79}
can be cast in the mean-field approximation (MFA) to be equal to
the internal energy of the system \cite{lang:etal:jpcm:00}:
\begin{equation}
F_{\rm ex}[\rho] \cong
{{1}\over{2}} \int\int {\rm d}^3 r {\rm d}^3 r' v(|{\bf r} - {\bf r'}|)
\rho({\bf r}) \rho({\bf r'}),
\label{dft.mfa}
\end{equation}
with the approximation becoming more accurate with increasing density.
Then, Eq.\ (\ref{dft.mfa}) immediately implies that in this limit
the direct correlation function $c(r)$ of the system,
defined as \cite{evans:79}
\begin{equation}
c(|{\bf r} - {\bf r'}|;\rho) =
-\lim_{\rho({\bf r}) \to \rho}
{{\delta^{2} \beta F_{\rm ex}[\rho({\bf r})]}\over
 {\delta \rho({\bf r}) \delta \rho({\bf r'})}},
\label{dcf.dft}                                                         
\end{equation}
becomes independent of the density and is simply proportional
to the interaction, namely
\begin{equation}
c(r) = -\beta v(r).
\label{mfa}
\end{equation}
Using the last equation, together with the Ornstein-Zernike
relation \cite{hansen:mcdonald}, we readily obtain an analytic expression
for the structure factor $S(k)$ of the system as
\begin{equation}
S(k) = {{1}\over{1 + \bar\rho t^{-1} \tilde \phi(k\sigma)}},
\label{sofq.analytic}
\end{equation} 
where 
$\tilde \phi(q) = \int {\rm d}^3x\exp(-{\rm i}{\bf q}\cdot{\bf x})\phi(x)$ 
is the Fourier transform of $\phi(x)$. 

Bounded and positive-definite interactions have been
studied in the late 1970s by
Grewe and Klein \cite{grewe:klein:jmpa:77,grewe:klein:jmpb:77}.
The authors considered 
a Kac potential
of the form:
\begin{equation}
v(r) = \gamma^d\psi(\gamma r),
\label{kac}
\end{equation}
where $d$ is the dimension of the space and $\gamma \geq 0$ is a
parameter controlling the range {\it and} strength of the potential.
Moreover, $\psi(x)$ is a nonnegative, bounded and integrable function.
Grewe and Klein showed rigorously that at the limit
$\gamma \to 0$, the direct correlation function of a system interacting
by means of the potential (\ref{kac}) is given by Eq.\ (\ref{mfa})
above. The connection with the case we are discussing here is
straightforward: as there are no hard cores in the system or a 
lattice constant to impose a length scale, the only relevant
length is set by the density and is
equal to $\rho^{-1/3}$ in our model and by
the parameter $\gamma^{-1}$ in model (\ref{kac}). In this respect,
the limit $\gamma \to 0$ in the Kac model of Grewe and Klein
is equivalent to the limit $\rho \to \infty$ considered here.

Although the limit of Grewe and Klein corresponds to $t \to \infty$ 
and $\bar\rho \to \infty$, the relation (\ref{mfa}) has been
shown to be an excellent approximation at arbitrarily low temperatures for high
enough densities \cite{lang:etal:jpcm:00}
and for temperatures
$t \gtrsim 1$ practically at {\it all densities} \cite{likos:etal:pre:01}.
Hence, the mean-field approximation is valid in a vast range of the
thermodynamic space of such systems, which has led to their characterisation
as {\it mean-field fluids} (MFF) \cite{ard:faraday:01}. Associated with the
structural relation (\ref{mfa}) are scaling relations of thermodynamic
quantities, arising from the compressibility sum rule \cite{hansen:mcdonald}:
\begin{equation}
f''(\rho) = -\tilde c(k = 0;\rho) = -4\pi\int_0^{\infty} r^2c(r;\rho)\,{\rm d}r,
\label{compress.sum.rule}
\end{equation}
where $f(\rho) = \beta F_{\rm ex}(\rho)/V$, and the primes denote the second
derivative. From Eqs.\ (\ref{mfa}) and (\ref{compress.sum.rule}) it then
follows
\begin{equation}
f(\rho) =
\frac{\beta\tilde v(0)}{2}\rho^2,
\label{mfa.fren}
\end{equation} 
with $\tilde v(0) = 4\pi\int_0^{\infty} r^2 v(r)\,{\rm d}r$. 
This simple scaling is not at all
equivalent to a second virial theory.  In fact, 
simple virial
expansions have a rather small radius of convergence for mean field
fluids \cite{ard:mft:00, ard:faraday:01}.
It then follows
that the excess pressure, chemical potential and compressibility satisfy
the scaling relations $P_{\rm ex} \sim \rho^2$, $\mu_{\rm ex} \sim \rho$
and 
$\chi_{\rm ex} \sim \rho^{-2}$ 
\cite{ard:mft:00, lang:etal:jpcm:00}.\footnote{Note that for polymers
in a good solvent $P_{\rm ex} \sim \rho^{3 \nu/(3 \nu -1)} \approx
\rho^{2.3}$ in the semi-dilute regime.  The density dependence of the
pair-potentials is necessary to properly describe this correction to
simple MFF behaviour \cite{bolhuis:jcp:00}.  But this in turn
implies that the extra factor $0.3$ in the scaling of the pressure
arises from the many-body interactions, since these are what cause the
density-dependence in the first place \cite{bolhuis:pre:01}.}
All these stem from the validity of the strong {\it structural}-MFA
relation (\ref{mfa}) which guarantees the validity of the weaker
{\it thermodynamical} MFA relation (\ref{mfa.fren}). In what follows,
we will argue that many ultrasoft systems can still satisfy the
thermodynamic relation (\ref{mfa.fren}) approximately, {\it without}
satisfying Eq.\ (\ref{mfa}). To distinguish between the two
classes, we now qualify the term ``mean-field fluids'' and call the
ones for which both the structural and thermodynamic MFA work well
{\it strong mean-field fluids}.  For such systems, for which the
potentials are typically also bounded, the density functional of 
Eq.\ (\ref{dft.mfa}) has been extended to mixtures \cite{ard:mft:00}
allowing a straightforward and transparent analysis of the phase
separation in the bulk, interfacial \cite{archer:evans:pre:01} and
wetting properties \cite{archer:evans:jpcm:02} of such mixtures.

\begin{figure}[hbt]
   \begin{center}
   \begin{minipage}[t]{6.5cm}
   \includegraphics[width=6.0cm,height=6.1cm]
   {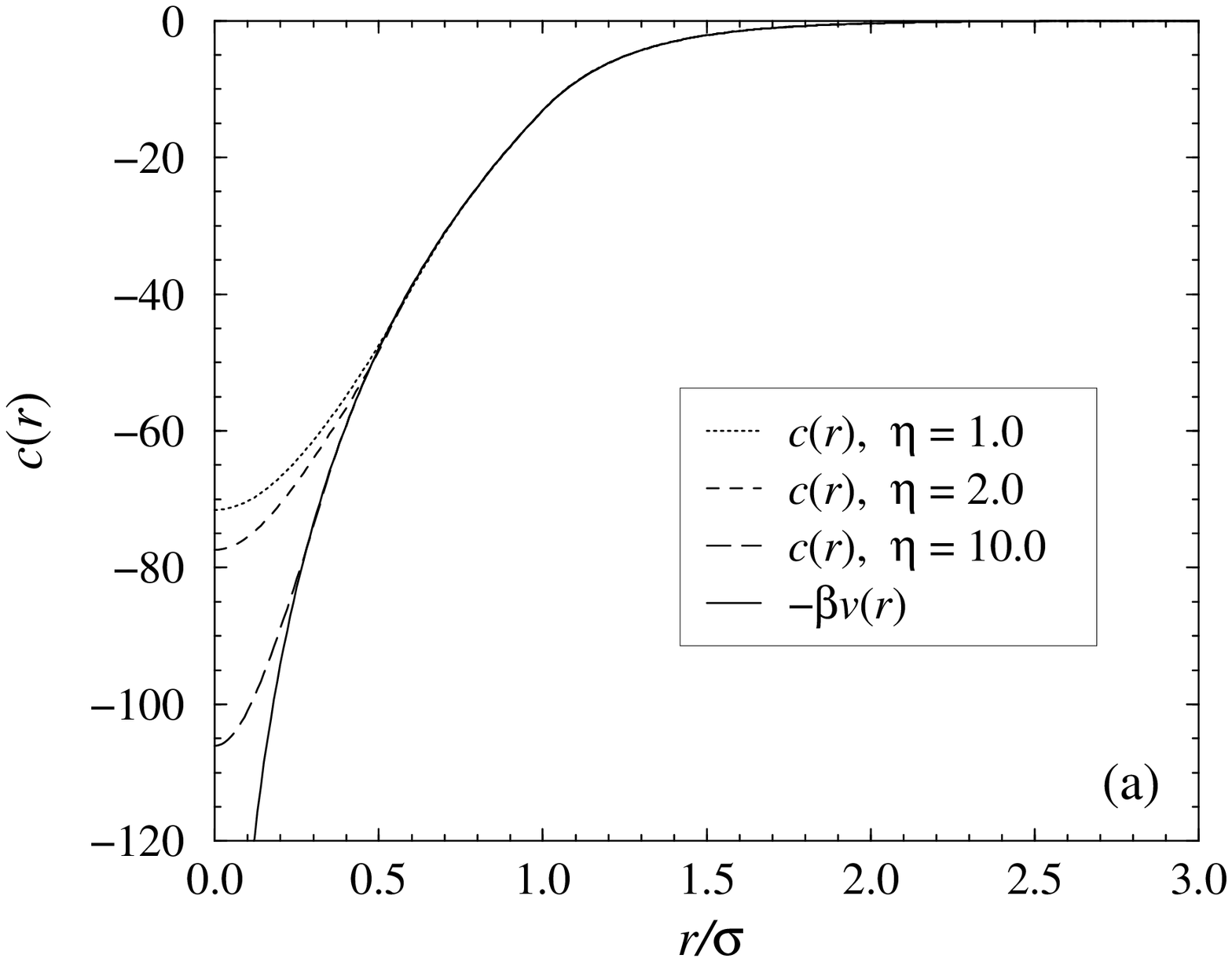}  
   \end{minipage}
   \begin{minipage}[t]{6.5cm}
   \includegraphics[width=6.0cm,height=6.1cm]
   {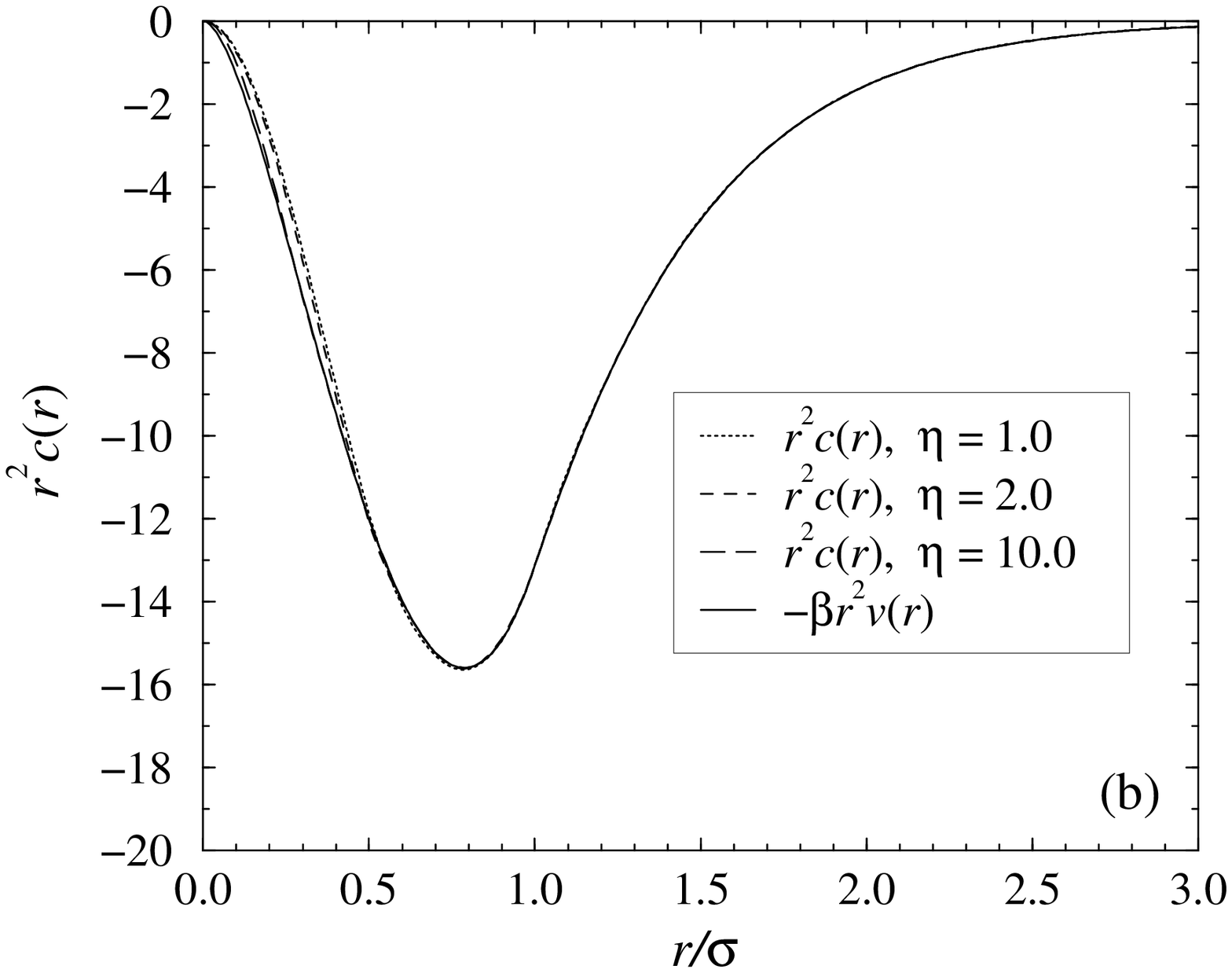}
   \end{minipage}
   \end{center}
   \caption{(a) Comparison between the direct correlation function
            of a $f = 32$ star polymer fluid at various densities, 
            obtained by solving the Rogers-Young closure, with
            the mean-field result, $-\beta v(r)$. (b) Same as in (a)
            but for the quantities $r^2 c(r)$ and $-\beta r^2 v(r)$.}
\label{cofr.plot}
\end{figure}

Let us now then turn our attention to 
interaction potentials such as those given in 
Eqs.\ (\ref{pot_ss2}) and (\ref{pot_ss}). These diverge at the
origin slowly enough, so that the three-dimensional integral
\begin{equation} 
\int{\rm d}^3r\, v(r) = 4\pi \int_0^{\infty} r^2 v(r)\,{\rm d}r\ =
\tilde v(0),
\label{fourtr}
\end{equation}
is finite and equal (by definition) to the value of the Fourier transform
of the potential $\tilde v(k)$ at $k =0$. It is now impossible
to satisfy
the strong mean-field condition of Eq.\ (\ref{mfa}) everywhere. Indeed, the
direct correlation function $c(r)$ has to remain finite at $r = 0$,
whereas the pair potential diverges.
Hence, as shown in Fig.\ \ref{cofr.plot}(a),
there will always exist a region in the neighborhood of the origin
in which Eq.\ (\ref{mfa}) is violated. At the same time,
it can be seen in this figure that the extent of this region shrinks
with increasing density, hence the fluid becomes more `strong mean-field'-like
as it gets denser. The
smaller $\tilde v(0)$, the lower the density at which the MFA for
$S(k)$, Eq.\ (\ref{sofq.analytic}), becomes a reasonable approximation.

\begin{figure}[hbt]
      \begin{center}
      \includegraphics[width=7.0cm,height=7.0cm]
      {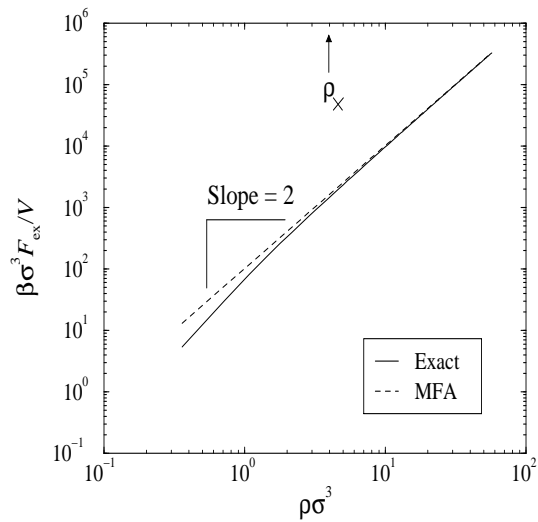}
      \end{center}
   \caption{Comparison of the mean-field approximation (dashed line)
            with the exact result (solid line) concerning the excess
            free energy density of the $f=32$ star fluid. The slope
            of the straight line is 2, indicating the quadratic
            dependence of the excess free energy density on particle
            density. The arrow indicates the location of
            the crossover density $\rho_{\times}$, above which the
            scaling of Eq.\ (\ref{mfa.fren}) holds with a relative
            error of less than $10\%.$}
   \label{fxliq.plot}
\end{figure}

The discrepancies between $c(r)$ and $-\beta v(r)$ become 
much less important when we
turn our attention to the thermodynamics. To obtain
the excess Helmholtz free energy, one needs only the {\it integral}
of $r^2 c(r)$, see Eq.\ (\ref{compress.sum.rule}). As demonstrated
in Fig.\ \ref{cofr.plot}(b), upon multiplication with the geometrical
factor $r^2$, the deviations of $c(r)$ from $-\beta v(r)$
become suppressed, so that we can write, to a very good approximation:
\begin{equation}
\int_0^{\infty}{\rm d}r\, r^2c(r;\rho) \cong
-\int_0^{\infty}{\rm d}r\, r^2\beta v(r).
\label{appr.integral}
\end{equation}
Eq.\ (\ref{appr.integral}) together with Eq.\ (\ref{compress.sum.rule})
yield an approximate scaling of the excess free energy of the
weak mean-field fluids with density that is identical to that
of the strong mean-field fluids, Eq.\ (\ref{mfa.fren}). The
accuracy of the approximation for the star polymer fluid 
with $f=32$ arms [Eq.\ (\ref{pot_ss})] is
shown in Fig.\ \ref{fxliq.plot}. The line labeled as exact free
energy there was obtained by solving the 
Rogers-Young closure \cite{rogers:young}
for the fluid at a wide density range and subsequently utilizing
the compressibility sum rule [Eq.\ (\ref{compress.sum.rule})] to
obtain the excess free energy. Comparisons with
simulations \cite{watzlawek:etal:jpcm:98}
have indeed demonstrated that this procedure delivers an
essentially exact numerical result. 

Clearly, the mean-field approximation improves with increasing
density, as the number of particles effectively interacting with
one another grows. The crossover density
$\rho_{\times}$ above which the quadratic scaling
of the free energy holds is $f$-dependent and grows with increasing $f$.
Indeed, the functionality acts as a prefactor that controls the
strength of the logarithmic divergence of the potential at the origin.
Formally, the mean-field approximation also becomes better with growing
spatial dimension $d$, as the geometrical prefactor $r^{d-1}$
multiplying $c(r)$ and $-\beta v(r)$ suppresses the small-$r$
discrepancies of the two more efficiently. We call systems for which
the mean-field idea holds only for the
thermodynamics {\it weak mean-field fluids}. 
If one na{\"\i}vely applies the {\it strong} 
mean-field relation, Eq.\ (\ref{mfa}),
to {\it weak} mean-field fluids, one obtains results for the 
structure factor $S(k) = [1-\rho\tilde c(k)]^{-1}$ that 
can be seriously
in error for finite $k$-values. {\it Only} at $k = 0$ and at sufficiently
high densities is it a reasonable approximation to set
$S(0) = [1+\rho\beta\tilde v(0)]^{-1}$.

\section{Clustering and reentrant melting}
\label{cluster.section}

In this section, we turn our attention to the phase behaviour of 
strong mean-field fluids. Two representatives of this class whose
phase behaviour has been studied in detail are the Gaussian core model 
(GCM) of
Eq.\ (\ref{vcom.eq}) and the `penetrable sphere model' (PSM) characterised
by the interaction potential $v_{\rm psm}(r) = \varepsilon\Theta(\sigma - r)$,
with the Heaviside step function $\Theta(x)$. Clearly, the PSM reduces
to the hard sphere model for $t = 0$. 

The GCM has been the subject of extensive investigations 
by Stillinger {\it et al.}\ in the 
late 1970s \cite{stillinger:76, stillinger:weber:78, stillinger:jcp:79,
stillinger:prb:79, stillinger:weber:80}. The $t = 0$ phase
diagram of the model was calculated, showing the existence of two
stable crystal structures, fcc for low densities and bcc for high ones.
In addition, a host of mathematical relations for the GCM has been
established and on the basis of free energy estimates is has been
postulated that the system displays reentrant melting behaviour at low
temperatures. On the basis of simulation studies at selected 
thermodynamic points, a rough phase diagram of the GCM has been
drawn \cite{stillinger:stillinger:97}. A detailed study of the
structural and phase behaviour of the GCM was carried out recently
by Lang {\it et al.} \cite{lang:etal:jpcm:00}. There, it was indeed
shown that for temperatures $t > 0.01$ the system remains fluid
at all densities, whereas for $t \leq 0.01$ reentrant melting is observed:
increasing the density, the system first undergoes a fluid $\to$ fcc
transition, followed by a structural fcc $\to$ bcc transition and
at higher densities the bcc solid remelts, i.e., a bcc $\to$ fluid
transition takes place. The width of the solid-phase region grows
with decreasing temperature. A structural signature of this unusual
phase diagram in the fluid phase above $t = 0.01$ is an anomaly in the
behaviour of the liquid structure factor $S(k)$. The height of its main peak 
first grows with increasing density and after achieving a maximum, it
decreases again, reflecting the stability of the fluid beyond the
reentrant melting. The Hansen-Verlet freezing
criterion \cite{hansen:verlet:69, hansen:schiff:73}
was shown to be satisfied at {\it both} the freezing and the reentrant 
melting lines \cite{lang:etal:jpcm:00}. Moreover, it was found that
at high densities not only the mean-field relation, Eq.\ (\ref{mfa})
is satisfied to excellent accuracy but also that the hypernetted
chain closure (HNC) becomes quasi-exact \cite{ard:mft:00, lang:etal:jpcm:00}.
The system becomes `quasi-ideal' at those densities, meaning 
that the radial distribution function has the limiting behaviour
$g(r) \to 1$. This, in conjunction with the mean-field property
$c(r) = -\beta v(r)$ and the exact relation 
$g(r) = \exp[-\beta v(r) -c(r) + g(r) - 1 +B(r)]$, forces the 
bridge function to obey the limit $B(r) \to 0$, hence rendering
the HNC exact.

The PSM was first studied in detail by 
by means of integral equation
theories, computer simulations and cell-model calculations at
small temperatures, $t \leq 0.3$ \cite{likos:psm:98}.
In contrast to the GCM, no
reentrant melting was found. Instead, the freezing line of the
system appeared to persist at all temperatures, and cascades of
clustering transitions in the solid were found, in which solids
with multiple site occupancies are stable with increasing temperature
and density. These findings were independently confirmed in 
a density-functional study of the 
low-temperature phase behaviour of the PSM \cite{schmidt:cecam:99}.
Sophisticated integral-equation approaches at arbitrarily high
temperatures revealed a loss of the solution along the `diagonal'
$t = \eta$ of the phase diagram \cite{fernaud:jcp:00}, again 
a feature pointing to an instability of the liquid by increasing
density at arbitrarily high temperatures. Thus, the PSM and the GCM
show completely different phase behaviours, although they are both
bounded and non-attractive potentials. There is a cascade of
clustering transitions for the former, enabling freezing at all
temperatures, and a reentrant melting for the latter, associated with
the inability to stabilise crystals above a certain critical 
temperature.\footnote{The use of the terminology `critical temperature'
here refers simply to the fact that above the said temperature
freezing is impossible and should not be confused with its standard
meaning in the realm of critical phenomena. There are no diverging
thermodynamic quantities here, no universality and all free energies
remain analytical functions in the neighborhood of the critical temperature.}
 
The key in understanding these two very different types of behaviour
lies in the strong mean-field character of these fluids and the
associated expression for the fluid structure factor, 
Eq.\ (\ref{sofq.analytic}). If the Fourier transform of the pair
interaction $\tilde \phi(k)$ has oscillatory behaviour 
(i.e., if $\tilde\phi(k)$ becomes negative for some $k$-values),
then at the wavenumber $k_{*}\sigma$ where 
$\tilde\phi(k\sigma)$ attains its
most negative value, $-|\tilde \phi(k_*\sigma)|$,
the liquid structure factor $S(k_{*})$ will display
a maximum. For any given temperature $t$, there exists then a
density $\bar\rho_{\rm s}(t)$
such that $\bar\rho_{\rm s}|\tilde \phi(k_*\sigma)| = t$,
causing a divergence of the fluid structure factor at $k_*$ and marking
a `spinodal line' at this finite wavenumber. Thus, the fluid cannot
be stable at all densities. As a matter of fact, freezing will take
place before the spinodal line is reached. The PSM clearly
belongs to this category, since the abrupt jump of the
pair interaction $v_{\rm psm}(r)$ at $r = \sigma$ causes 
long-range oscillations of the potential in Fourier space. 
By employing the 
Hansen-Verlet criterion, $S(k_*) \cong 3$, it has been 
found \cite{likos:etal:pre:01} that in the
PSM freezing takes place at the 
`diagonal' $t = \eta$ on
which the integral equation approach of Fernaud 
{\it et al.}\ \cite{fernaud:jcp:00} breaks down. If, on the
other hand, the Fourier transform of the potential, $\tilde\phi(k)$
is a positive-definite, monotonically decreasing function of $k$,
Eq.\ (\ref{sofq.analytic}) assures that
at sufficiently high temperatures, where the mean-field approximation
is valid at all densities \cite{likos:etal:pre:01}, $S(k)$ is a 
monotonic function of $k$ approaching rapidly the value $S(k) = 1$
with increasing $k$ and being deprived of any peaks. The lack of
peaks in the structure factor implies the lack of any
tendency within the liquid towards spontaneous formation of spatially
modulated patterns. Thus, the fluid remains stable for all densities
at sufficiently high temperatures. This, combined with the observation 
that at low temperatures and densities bounded potentials all become
hard-sphere like and hence they must cause a freezing transition there,
leads to a reentrant-melting scenario for such systems. Clearly,
the GCM belongs to this category. Representative results for a particular
family of strong mean-field fluids and schematic phase diagrams 
can be found in Ref.\ \cite{likos:etal:pre:01}. 

\section{Exotic crystal phases}
\label{exotic.section}

We now focus on weak-mean field fluids, for which no
simple criterion for their freezing behaviour can be established,
since Eqs.\ (\ref{mfa}) and (\ref{sofq.analytic}) are not satisfied
any more. The star-polymer fluid characterised by the pair
potential of Eq.\ (\ref{pot_ss}) is a case in 
point.\footnote{We consider here the case $f > 10$, for which
indeed the interaction of Eq.\ (\ref{pot_ss}) holds, and not
the low-functionality case for which the interaction 
of Eq.\ (\ref{pot_ss2}) is valid. The reason is that at low
functionalities the stars do not freeze at any density and hence
they are not an appropriate system for considering 
thermodynamically stable crystals.} The physical system of
star polymers provides an excellent testbed for the investigation
of the thermodynamic stability of more complicated crystals than 
usual fcc- and bcc-lattice arrangements. 

The fcc-lattice is the
one favoured by hard interactions, since it has the property of
maximising the available volume and hence the entropy
of the particles for a given particle density \cite{hales:00}.
On the other hand, the presence of `soft tails' in the potential
has the effect of favouring the more open bcc-lattice, as was
convincingly demonstrated for the case of the screened Coulomb
potential (Yukawa interaction) arising in charge-stabilised
colloidal 
suspensions \cite{hone:83, kremer:robbins:grest:86, robbins:kremer:grest:88, sirota:89}.
These two common lattices were considered for a long time to be the
only `candidates' in a search for stable crystals for given interatomic
potentials. However, in modern colloidal science, new possibilities open
up. It is technically possible to manufacture micelle-like particles
featuring a hard core and a soft, fluffy corona of grafted or adsorbed
polymer chains, with the thickness $L$ of the latter being much larger than
the radius $R_c$ of the former. Star polymers correspond to the case
$R_c \ll L$; the theoretical arguments leading to the effective potential
of Eq.\ (\ref{pot_ss}) are in fact based on the assumption $R_c \to 0$. 
Under these physical circumstances, the effective interaction between the
micelles is dominated by the ultrasoft repulsion between the overlapping,
flexible coronas and {\it not} by the excluded volume interactions of the
hard cores. Thereby, the requirement of maximising of the volume available
to each particle does not play the decisive role any more. 
 
\begin{figure}[hbt]
      \begin{center}
      \includegraphics[width=10.0cm]
      {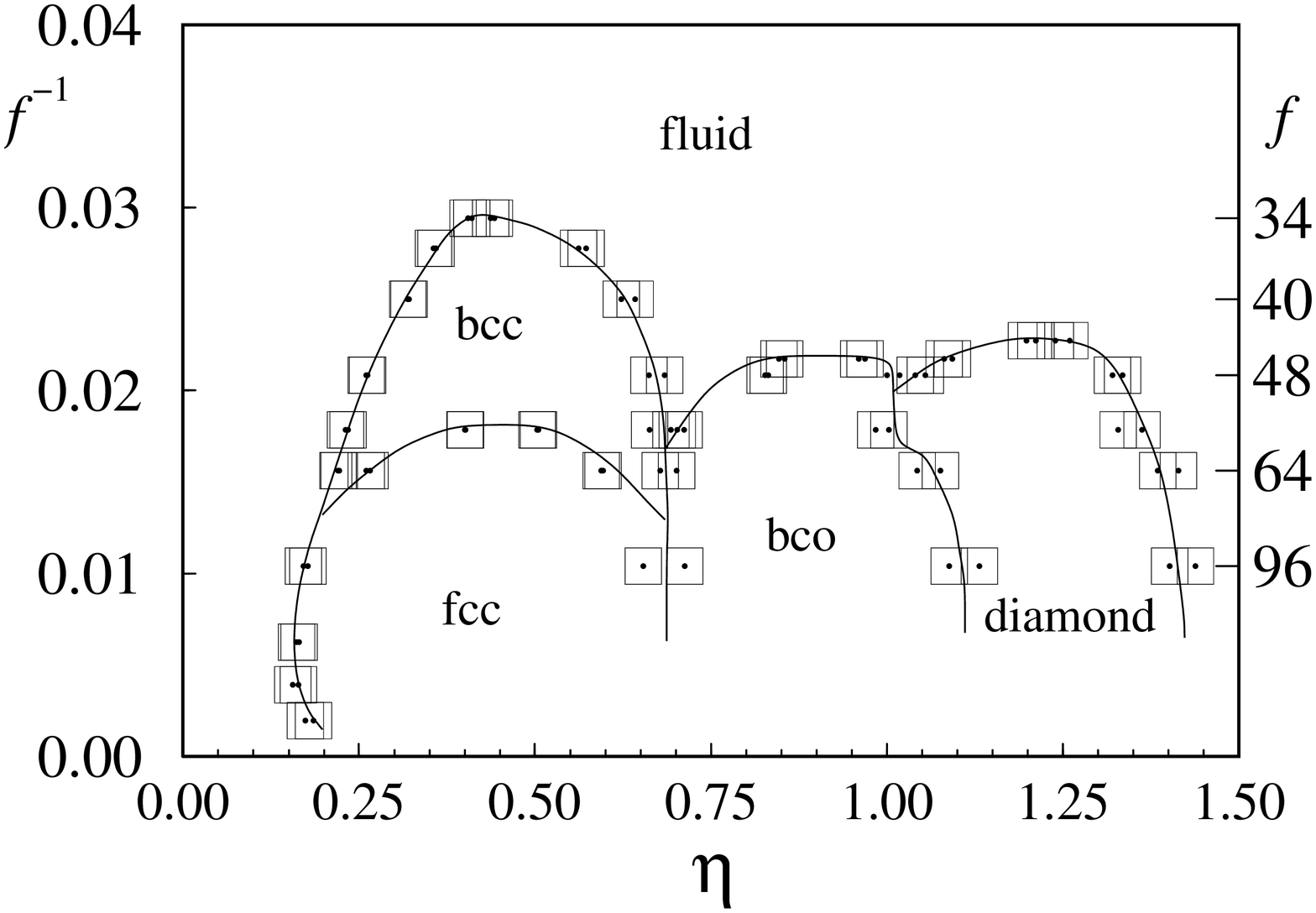} 
      \end{center}
   \caption{The phase diagram of star polymers. 
            The symbols denote 
            simulation results for the pairs of coexisting densities
            and the lines are guides to the eye. Notice that the 
            density gaps at the phase boundaries are very narrow.
            (Redrawn from Ref.\ \cite{martin:prl:99}.)}
   \label{stars.phdg.fig}
\end{figure}

These properties manifest themselves in the phase diagram of star-polymer
solutions. Due to the irrelevance of
the temperature for this entropic interactions,
the phase diagram was drawn in the 
($f,\eta$)-plane by 
Watzlawek {\it et al.}\ \cite{martin:prl:99, martin:phd:00},
where $\eta = \pi\rho\sigma^3/6$ and $\rho = N/V$ is the
number density of $N$ stars in the volume $V$. The
phase diagram is shown in
Fig.\ \ref{stars.phdg.fig}. The fluid phase remains stable at
all concentrations for $f < f_c = 34$, a result that confirms
and makes precise early scaling-argument predictions of 
Witten {\it et al.}\ \cite{witten:pincus:86:1}. For $f > f_c$
and at packing fractions $0.15 \lesssim \eta \lesssim 0.70$,
the usual fcc- and bcc-crystals are seen to be stable, the
former for larger and the latter for smaller functionalities.
This is consistent with the fact that the effective potential
of Eq.\ (\ref{pot_ss}) has a Yukawa decay length scaling as
$1/\sqrt{f}$, hence large $f$ is analogous to the strongly
screened charge-stabilised colloidal suspensions. However,
for $\eta \gtrsim 0.70$, unusual crystal structures appear.
First, in the domain $0.70 \lesssim \eta \lesssim 1.10$, a
body-center-orthogonal (bco) crystal is thermodynamically
stable. The bco-lattice is characterised by a body-centered,
orthogonal conventional unit cell with three unequal sides
and reduces to the bcc-lattice for ratios $1:1:1$ between the
sides and to the fcc for ratios $1:1/\sqrt{2}:1/\sqrt{2}$
\cite{ashcroft:mermin}. The bco-lattices appearing in this
region of the phase diagram feature strongly anisotropic
unit cells with typical size ratios $1:0.6:0.3$. Thus, these
are crystals with coordination number 2. For packing fractions
$1.10 \lesssim \eta \lesssim 1.50$, the diamond lattice with 
coordination number 4 turns out to be stable. Thus, we see
that {\it very open} structures with their characteristically low
coordination numbers are stabilised by the ultrasoft star-star
potential. This feature has been attributed to the very slow
divergence of the interaction as $r \to 0$, combined with its
crossover to a Yukawa-form for $r > \sigma$ 
\cite{likos:physrep:01, martin:prl:99, martin:phd:00}. 
Indeed, in such circumstances
it may be energetically preferable for the system to have a small
number of nearest neighbours at a small distance from any given
lattice point than a large number of neighbours at a greater
distance, as is the case for the optimally-packed fcc-lattice.

\begin{figure}[hbt]
   \begin{center}
   \begin{minipage}[t]{6.5cm}
   \includegraphics[width=6.0cm,height=6.1cm]
   {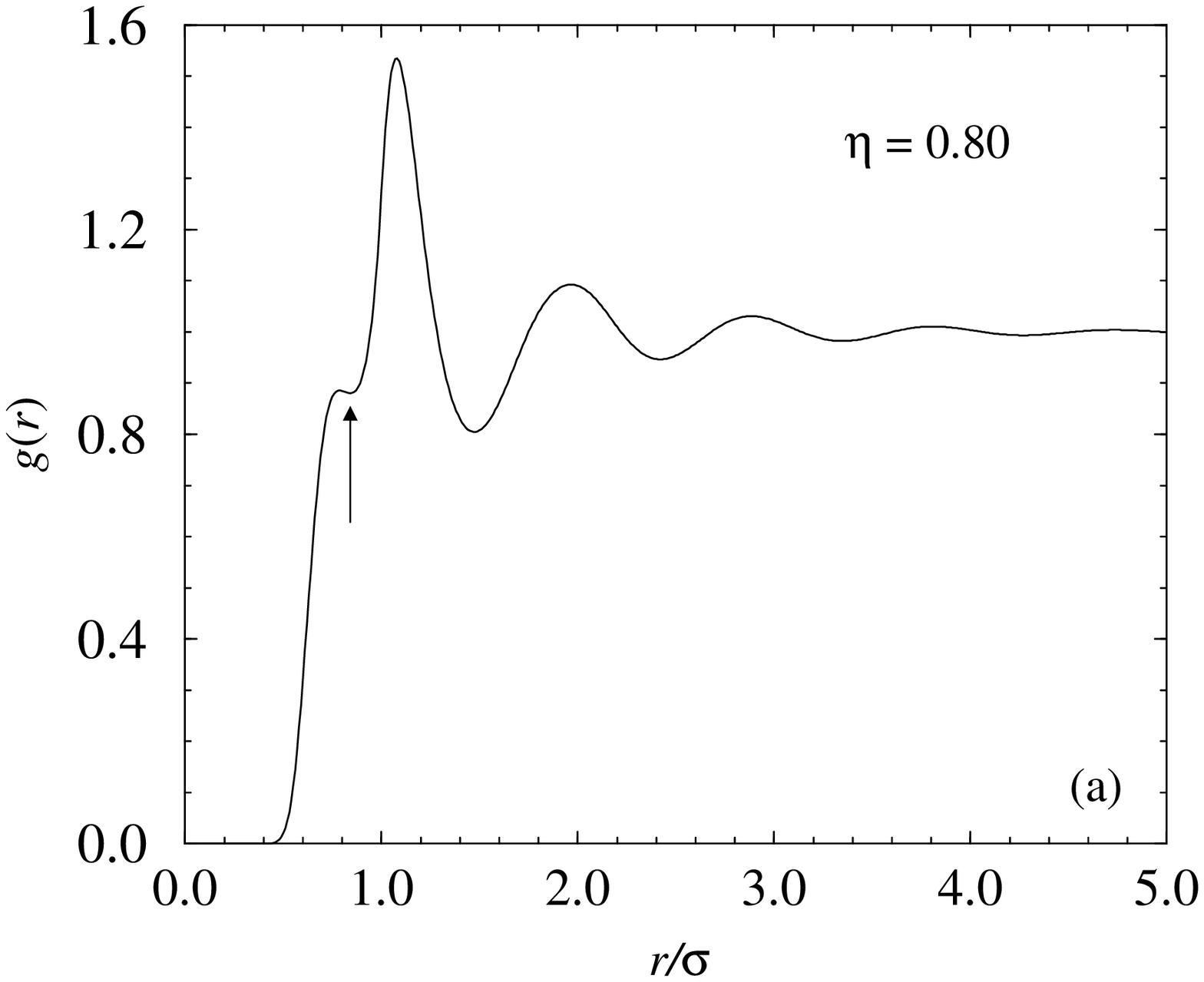}
   \end{minipage}
   \begin{minipage}[t]{6.5cm}
   \includegraphics[width=6.0cm,height=6.1cm]
   {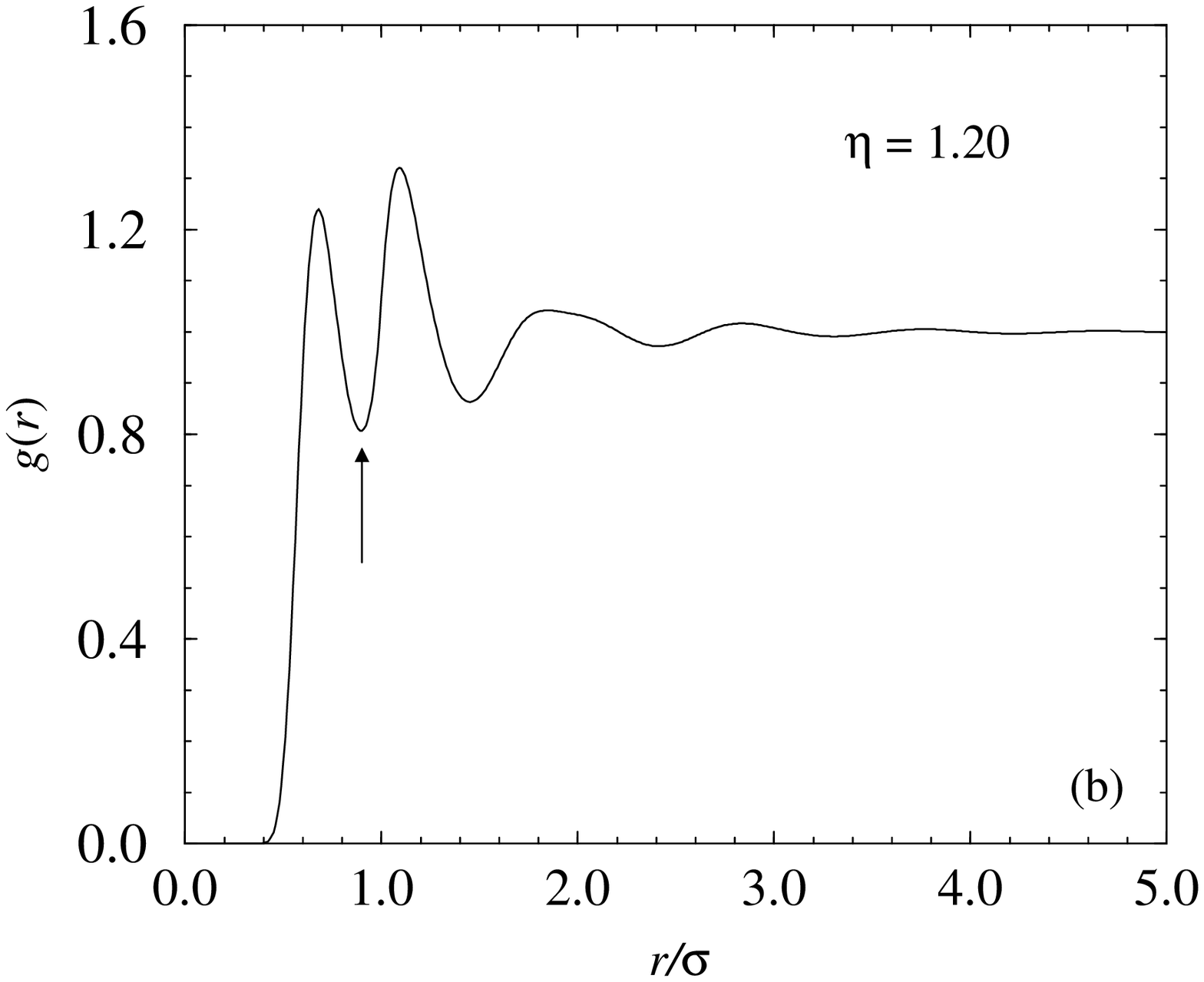}
   \end{minipage}
   \end{center}
   \caption{The radial distribution function of a star-polymer
            fluid of functionality $f=32$ at two different
            packing fractions, $\eta = 0.80$ [(a)] and 
            $\eta = 1.20$ [(b)]. The arrows indicate the positions
            $r_{\rm min}$ that define the borderline of the first 
            coordination shell in the fluid phase.}
\label{gofr.fig}
\end{figure}

In the fluid there exist clear structural signatures both for the
topology of the phase diagram of Fig.\ \ref{stars.phdg.fig} and for
the variety of the crystal phases featured there. The reentrant 
transition
from a fluid to a bcc-lattice and then again to a fluid, occurring
for $34 \lesssim f \lesssim 48$, is manifested in fluid structure
factors $S(q)$ that show a main peak that first grows with increasing
density and then drops again \cite{watzlawek:etal:jpcm:98},
as in the case of the Gaussian core model mentioned above. Moreover,
once more the Hansen-Verlet freezing 
criterion \cite{hansen:verlet:69, hansen:schiff:73} was found to
be satisfied on both sides of the freezing and reentrant melting
line. The radial distribution function $g(r)$ of the fluid at
various densities, on the other hand, carries the signature
of a local coordination that resembles that of the thermodynamically
neighbouring solids. To demonstrate this, we show in Fig.\ \ref{gofr.fig}
the function $g(r)$ for star polymer fluids at $f=32$, which are
thermodynamically stable, at packing fractions $\eta = 0.80$ and
$\eta = 1.20$. Comparison with Fig.\ \ref{stars.phdg.fig} shows
that the former corresponds to a state at the vicinity of the 
bco-phase and the latter to one at the vicinity of the diamond phase.
Consider now the average coordination number $z$ in the fluid phase,
defined as
\begin{equation}
z = 4\pi\rho\int_0^{r_{\rm min}} r^2g(r)\,{\rm d}r,
\label{z.eq}
\end{equation}
where $r_{\rm min}$ is the position for which $g(r)$ has its first
minimum and is indicated by the arrows in Fig.\ \ref{gofr.fig}. For
the two packing fractions shown we obtain $z = 1.95$
at $\eta = 0.80$
and $z = 4.03$ at $\eta = 1.20$. 
The first is very close to the coordination
number $z_{\rm bco} = 2$ of the neighboring bco-lattice and the
latter to $z_{\rm diam} = 4$ of the diamond lattice. The fluid
distribution functions contain local correlations that point to the
ordering of the incipient crystal phases.

The $g(r)$'s of the fluid above the overlap density,
$\eta \gtrsim 1.0$, show in addition anomalous behaviour featuring
two distinct length scales, as is clear from Fig.\ \ref{gofr.fig}(b).
As analysed in detail in Ref.\ \cite{watzlawek:etal:jpcm:98}, two
characteristics of the interaction are responsible for this behaviour:
on the one hand, the existence of the crossover of the interaction
of Eq.\ (\ref{pot_ss}) at $r = \sigma$ from a logarithmic to 
an exponentially decaying form. And on the other, the ultrasoftness
of the logarithmic potential, allowing the existence of fluids
at arbitrarily large  densities (for $f < f_c$), a feature unknown
for the usual interactions appearing in liquid-state physics and
which are all `perturbations' of the Hard-Sphere potential
(e.g., Lennard-Jones, inverse-powers etc.) Thus, ultrasoft potentials
carry unique structural signatures which should in principle be
visible in scattering experiments from soft, polymeric fluids.  

A great deal of insight into the general physical mechanisms driving the
stability of open structures in soft systems
was gained through the recent work of
Ziherl and Kamien \cite{ziherl:prl:00, ziherl:jpcb:01}. 
They considered in full generality 
systems with particles composed of a hard core and
long, deformable coronas and argued as follows. At any given density
above the overlap concentration, the coronas are forced
to overlap and compress, which gives rise to an entropic
free energy cost. The volume 
available to the coronas is fixed and equal to the difference
of the total volume minus that occupied by the hard cores.
Denoting by $d$ the thickness of the coronal layer and $A$ 
the total area of an imaginary membrane separating the compressed
coronas, it then turns out that the product $Ad$ is constant.
As the free energy cost for the compression of the chains
increases with decreasing thickness $d$, it turns out that
favourable phases are those for which the interfacial area
$A$ is minimal. Thereby, the problem reduces to that of determining
the ordered arrangement of point particles that generates Wigner-Seitz (WS)
cells having the smallest possible area for a given density. It is then 
conceivable that the fcc-lattice will be unfavoured, since
its WS cell has a larger area than that of the bcc,
for instance. In this way, Ziherl and Kamien established a beautiful
connection of this problem 
with Lord Kelvin's celebrated 
question of determining the area-minimising
partition of space for an arrangement of soap bubbles
of equal volume \cite{kelvin:1887}.

\begin{figure}[hbt]
      \begin{center}
      \includegraphics[width=7.0cm]
      {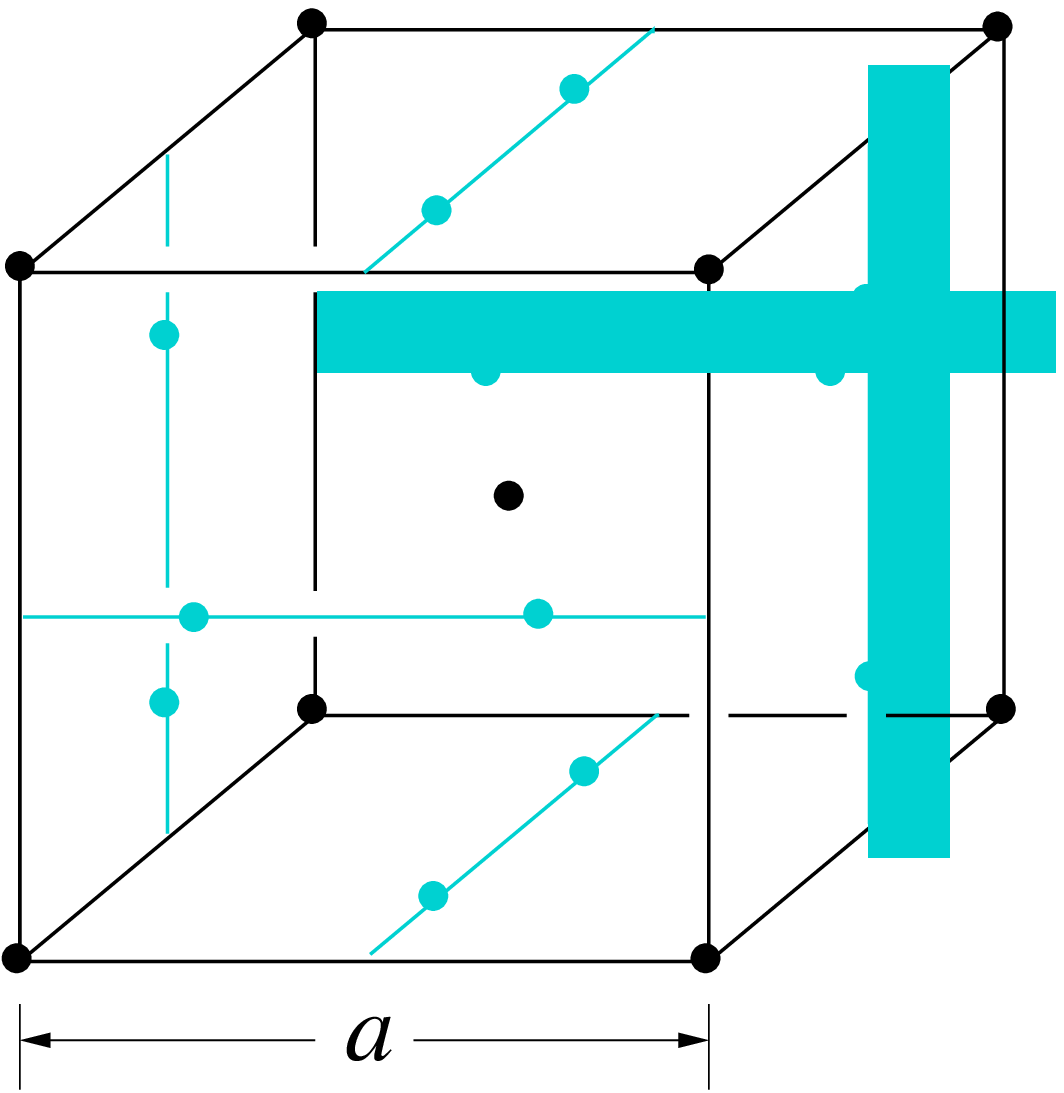}
      \end{center}
   \caption{The conventional unit cell of the A15-lattice.}
   \label{a15.fig}
\end{figure}

Following these arguments, it then turns out that there exists
yet another candidate phase that has an area even smaller than
the bcc-lattice \cite{weaire:94}, namely the A15-lattice \cite{rivier:94}.
Self-assembled micelles of dendritic molecules with a particular
architecture have been experimentally seen to crystallise into
this phase \cite{balag:97}. The conventional unit cell of
the A15-lattice is shown in Fig.\ \ref{a15.fig}. It can be thought
of as the cell of a bcc-lattice (dark points) decorated with 
`dimers' (light points) running along the middle 
of the faces. The dimers are oriented
parallel on opposite phases, forming thus columns through space.
The orientation of the dimers lying on intersecting faces is
perpendicular to one another, so that one third of all dimers lies
along each of the three Cartesian directions in space. The dimer
length is $a/2$, where $a$ is the edge length of the cube, and
it is placed symmetrically along the face, i.e., the distance of
any monomer to its nearest edge is $a/4$. The A15-lattice is
not a Bravais lattice; it can be constructed as a simple cubic (sc)
arrangement with an eight-member basis, thus it contains 8 sites
per conventional cell. Its WS-cell is a Goldberg 
decatetrahedron\footnote{We prefer the term {\it decatetrahedron}
for a polyhedron with 14 faces instead of the term
{\it tetrakaidecahedron}, often employed in the literature.
The latter, inspired from ancient Greek, literally means
`four-and-ten-faced polyhedron', whereas the former, consistent
with modern Greek, has the much more logical translation
`fourteen-faced polyhedron'. One of us (CNL) believes
that modern Greek language deserves a fair chance against its
classical predecessor, which has influenced scientific terminology
for quite some time.}
consisting of two hexagonal and 12 pentagonal
faces \cite{ziherl:jpcb:01}. 

Semi-quantitative calculations on the stability of the A15-lattice
were carried out by 
Ziherl and Kamien \cite{ziherl:jpcb:01}, 
using a model `square-shoulder' potential
within a simplified cell model. Narrow regions in thermodynamic
phase space were found, in which the A15-lattice was stable
but this finding is uncertain in view of the approximations
involved and the limited extent of this region. To investigate
this question in more detail, we have employed extensive lattice-sum
calculations for the star-polymer system, using the pair potential
of Eq.\ (\ref{pot_ss}) and extending both the set of candidate
lattices and the region of densities we looked at. We compared
between the sc, diamond, bco (which includes the fcc- and bcc-lattices
as special cases) and A15-lattices in the regions $0 \leq \eta \leq 2.50$
and $32 \leq f \leq 256$, at selected arm numbers $f$ to be shown
below. The lattice sums were performed by fixing the particles
at the prescribed lattice positions and keeping them frozen there,
i.e., no thermal fluctuations (harmonic corrections) were taken into account.
This approach reproduces very well the solid part
of the phase diagram of Fig.\ \ref{stars.phdg.fig}: the
free energy of the star-polymer crystals turns out to be dominated
by the lattice-sum term, the corrections to it from the 
particle oscillations
as well as the entropic contribution from the same playing only
a minor role. In this way, we are of course unable to compare
the solid free energies with those of the fluid, therefore no prediction
about melting can be made. However, for large enough $f$, the interaction is steep
enough, so that the system will be definitely be in a crystalline
phase for which the lattice sums provide an accurate prediction of the
most stable structure among the candidates.

For the bco-phases we minimised the lattice sums with respect 
to the two size ratios, $r_1 = b/a$ and $r_2 = c/a$ between the edge lengths,
$a$, $b$ and $c$ at any given density. Without loss of generality,
we assume in what follows that $a$ is the longest of the three
edges, thus $0 < r_1, r_2 \leq 1$. The minimised bco-energy
was then compared with the energies of all other lattices; the one
with the smallest lattice energy per particle wins. 

\begin{figure}[hbt]
      \begin{center}
      \includegraphics[width=10.0cm]
      {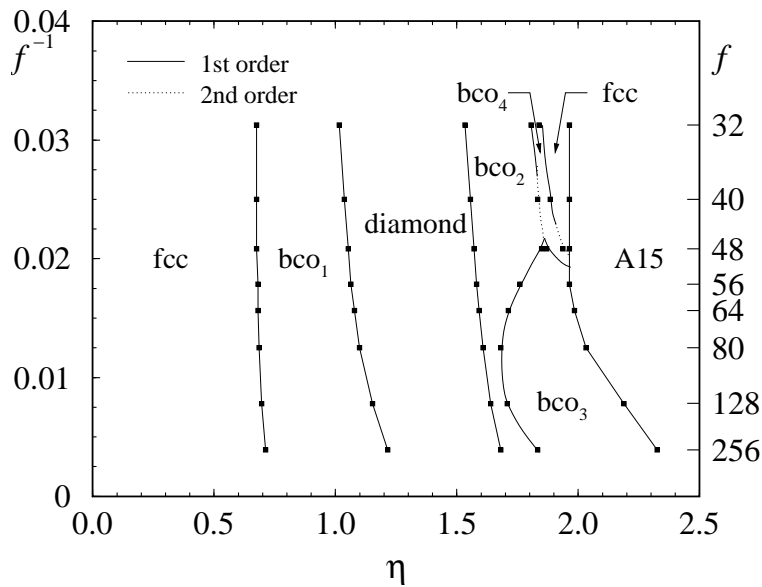}
      \end{center}
   \caption{The zero-temperature phase diagram of star polymer 
            solutions, obtained after the minimisation of lattice
            sums for particles interacting by means of the potential
            of Eq.\ (\ref{pot_ss}). The phase denoted bco$_1$
            in this figure is the same as the one denoted bco
            in Fig.\ \ref{stars.phdg.fig} but now it has to 
            be distinguished from the additional bco-phases showing
            up at higher densities and named bco$_2$, bco$_3$ and bco$_4$.}
   \label{latticesums.fig}
\end{figure}

The
`zero-temperature' phase diagram\footnote{`Zero-temperature' is just
a convenient way to refer to the assumption of frozen particles
at the lattice sites and does not refer to the {\it real} temperature
$T$ of the system. The latter is an irrelevant thermodynamic variable
because the entropic interaction of Eq.\ (\ref{pot_ss}) is proportional
to $k_BT$ and the thermal energy is the {\it only} energy
scale of the problem.}
obtained this way is shown
in Fig.\ \ref{latticesums.fig}.
First, we note that the phases being stable up to $\eta \cong 1.50$
are precisely those also seen in the finite-temperature phase diagram of 
Fig.\ \ref{stars.phdg.fig} and also that the phase boundaries 
based on the lattice sums agree very well with the ones at finite
temperatures. The A15-phase does not alter the hitherto explored
part of the phase diagram. However, for $\eta \gtrsim 1.50$, a host
of new phases and of transitions between those show up. The
A15-phase is stable at the high-density part of the phase diagram,
confirming thus explicitly the prediction of Refs.\ \cite{ziherl:prl:00}
and \cite{ziherl:jpcb:01} that this phase is a suitable candidate
at the high concentrations of ultrasoft particles. Nested between
the stability domain of the A15-lattice and the diamond lattice, 
four new bco-phases and (iso)structural 
transitions between those show up, having the following characteristics.

\begin{figure}[hbt]
   \begin{center}
   \begin{minipage}[t]{6.5cm}
   \includegraphics[width=6.0cm,height=6.1cm]
   {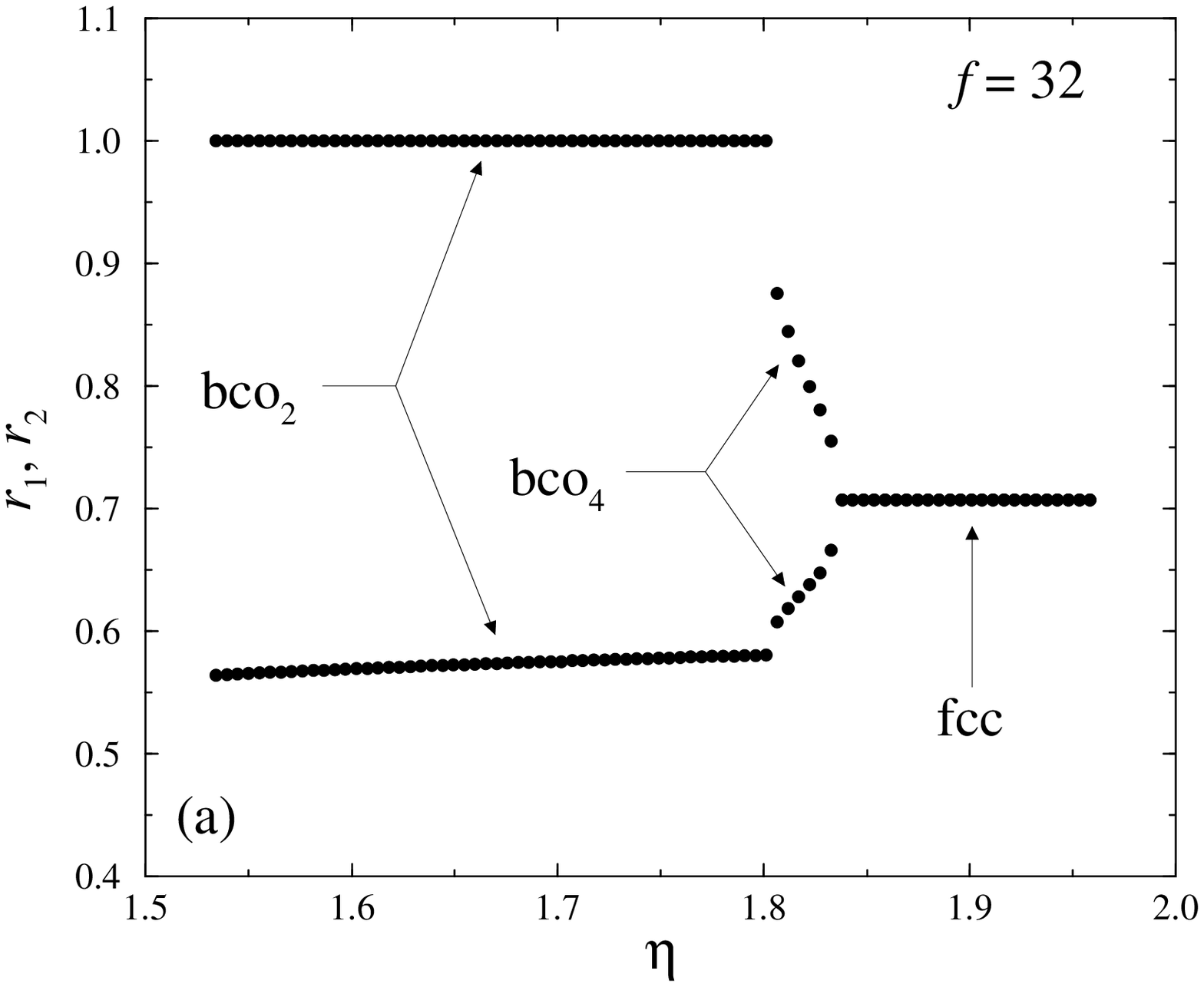}
   \end{minipage}
   \begin{minipage}[t]{6.5cm}
   \includegraphics[width=6.0cm,height=6.1cm]
   {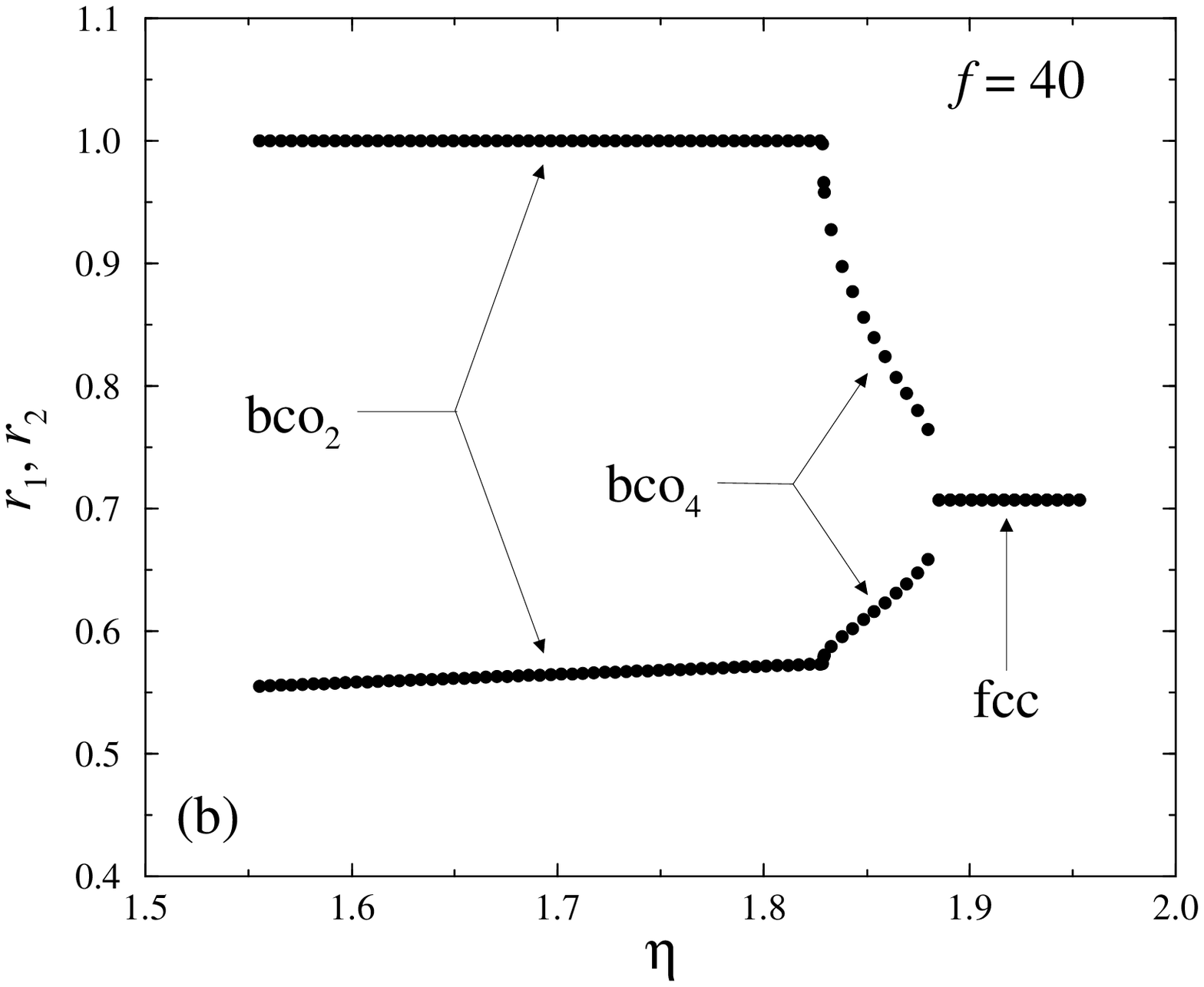}
   \end{minipage}
   \begin{center}
   \end{center}
   \begin{minipage}[t]{6.5cm}
   \includegraphics[width=6.0cm,height=6.1cm]
   {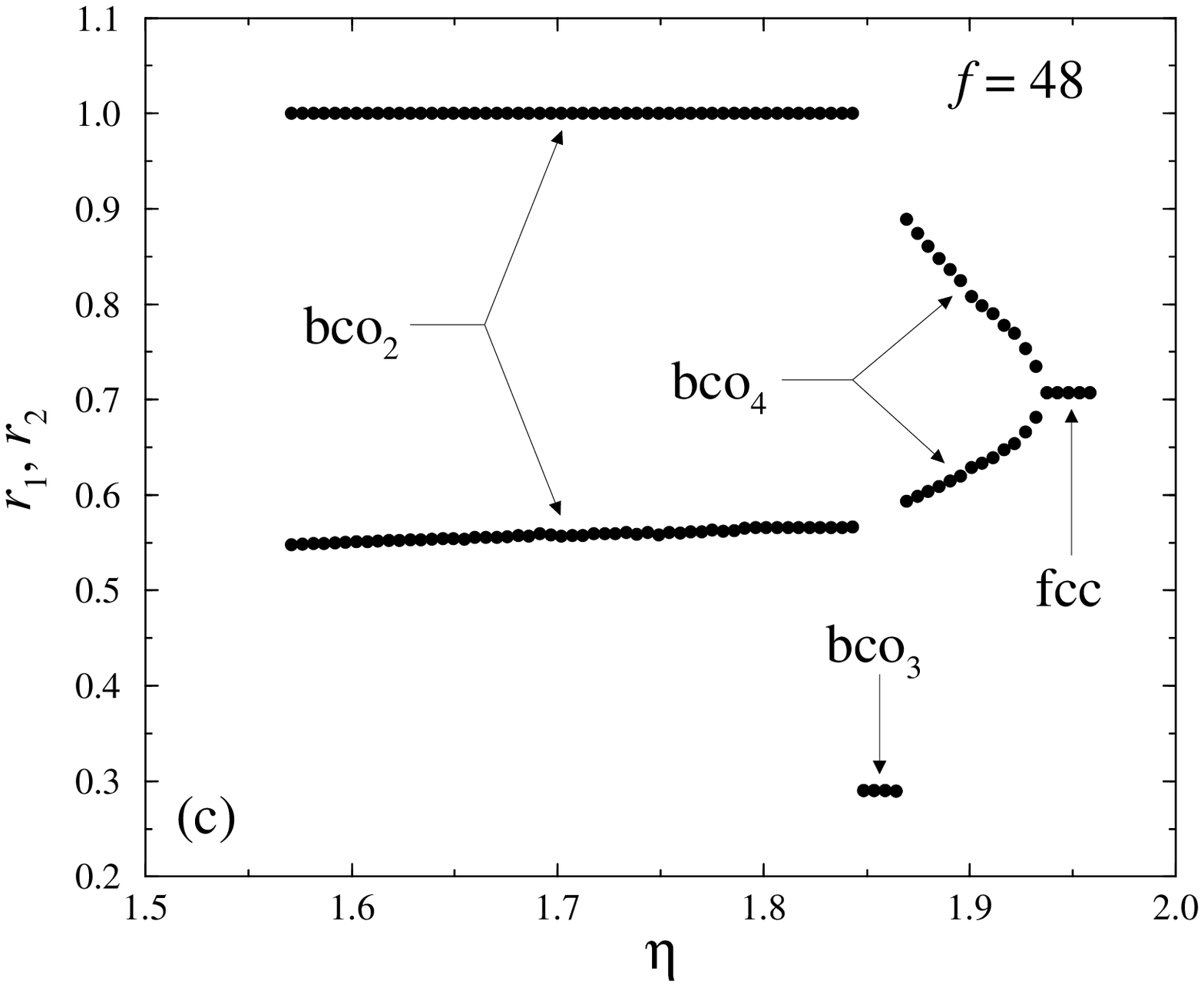}
   \end{minipage}
   \begin{minipage}[t]{6.5cm}
   \includegraphics[width=6.0cm,height=6.1cm]
   {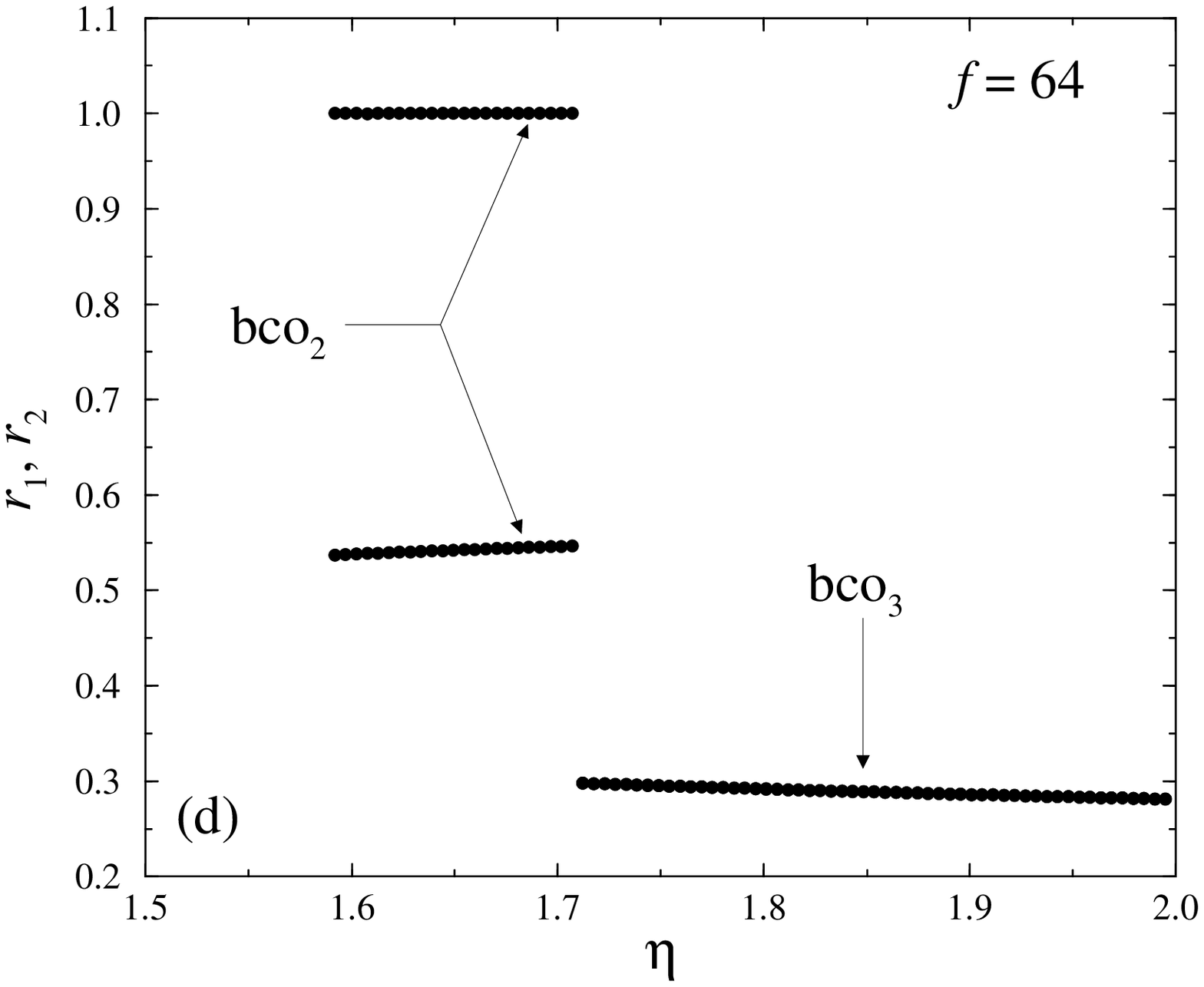}
   \end{minipage}
   \end{center}
   \caption{The optimal size ratios of the various bco-phases 
            occurring in the phase diagram of Fig.\ \ref{latticesums.fig}
            between the diamond and the A15-phase, within their domains
            of stability. Also shown is the 
            characterisation of those phases. The four different 
            functionalities are indicated on the plots.}
\label{ratios.fig}
\end{figure}

The phase denoted bco$_2$ has size ratios $r_1 = 1$ and $r_2 \cong 0.55$,
the former being constant at all densities and functionalities and
the latter showing very weak variation, see also Fig.\ \ref{ratios.fig}.
Hence, the unit cell of the bco$_2$-phase is anisotropic only in
one Cartesian direction and has a wide basis and height that is
smaller than the base edge length. Accordingly, the coordination
number of this phase is 2. The bco$_3$-phase, dominating at high
functionalities, has size ratios $r_1 = r_2 \cong 0.3$ which also
show very little variation with $\eta$ and $f$, see Fig.\ \ref{ratios.fig}.
Similarly to the bco$_2$-phase, therefore, the cell of the bco$_3$-phase
also has anisotropy in only one Cartesian direction. Contrary to it,
however, the height of the cell is now {\it larger} than the edge
length of the base and therefore the number of nearest neighbours is 4.
The phase transition bco$_2$ $\to$ bco$_3$ is first order:
as can be seen in Figs.\ \ref{ratios.fig}(c) and (d), the size ratios
jump abruptly from the values ($1, 0.55$) of bco$_2$ to the values 
($0.3, 0.3$) of bco$_3$. These are the only stable bco-phases in
this part of the phase diagram as long as $f \gtrsim 56$. 

For functionalities $32 \leq f \lesssim 56$, two more bco-phases show
up. First, there is a narrow region occupied by the bco$_4$-phase,
which has all three edge lengths of its unit cell different
and is thus similar to the bco$_1$-phase discovered before.
As seen in Figs.\ \ref{ratios.fig}(a), (b) and (c), these ratios
evolve
from the values $r_1 = 1$ and $r_2 \cong 0.55$ of the bco$_2$-phase
towards the values $r_1 = 1/\sqrt{2}$ and $r_2 = 1/\sqrt{2}$ of the
fcc-phase. 

The order of the
transitions between those phases has been investigated numerically.
Within the limits of accuracy of our numerical code, we have found that
for $f = 32$ and $f = 40$ 
the transition bco$_4$ $\to$ fcc is first order, i.e., the size ratios
of the bco$_4$-phase jump with density to the ratios
$r_1 = r_2 = 1/\sqrt{2}$ of the fcc-phase abruptly. This is
seen in Figs.\ \ref{ratios.fig}(a) and Fig.\ \ref{ratios.fig}(b).
For $f = 48$, the transition is second-order with the 
bco$_4$-size ratios evolving to the fcc-ones smoothly, see
Fig.\ \ref{ratios.fig}(c). 
The nature of the transition 
bco$_2$ $\to$ bco$_4$ is also mixed.
For $f = 32$, [Fig.\ \ref{ratios.fig}(a)], a finite 
jump of the values of the ratios was found, although the step size
in changing the packing fraction was made as small as $5 \times 10^{-5}$.
Thus, we characterise this transition at $f = 32$ as first-order.
For $f = 40$, [Fig.\ \ref{ratios.fig}(b)] this transition appears to
be second-order. Hence, we conclude that there must be a line of
second-order transitions terminating at a tricritical point 
between $f = 40$ and $f = 32$, (for the bco$_2$ $\to$ bco$_4$
transition) and similarly a tricritical point 
between $f = 48$ and $f = 40$ (for the bco$_4$ $\to$ 
fcc-transition) to be succeeded by lines of first-order
transitions.

All four bco-phases and the three transitions among them,
bco$_2$ $\to$ bco$_3$ $\to$ bco$_4$ $\to$ fcc, appear only in a very
narrow $f$-range around $f = 48$, see Fig.\ \ref{ratios.fig}(c). 
The line of second-order transitions bco$_2$ $\to$ bco$_4$ meets
the lines of first-order transitions bco$_2$ $\to$ bco$_3$ and
bco$_3$ $\to$ bco$_4$ at a critical endpoint located at $f$ slightly
less than 48. Similarly, the line of second-order transitions
bco$_4$ $\to$ fcc also terminates at a critical endpoint, meeting
the first-order lines bco$_3$ $\to$ A15 and fcc $\to$ A15. It is
an intriguing phenomenon that such a richness in the stable crystal
structures and in the nature of the transitions among them
occurs as the result of a simple, spherically symmetric interaction
and this points to the many surprises of ultrasoft potentials
and their tendency to produce open, exotic structures. At the
same time, it must be pointed out that we do not expect the
solid phases occurring for $f < 48$ to survive the competition
with a fluid. Indeed, as can be seen from Fig.\ \ref{stars.phdg.fig},
the critical functionality $f_c$ below which no solids are 
stable increases with increasing density. The bcc-phase is extinguished
by the fluid for $f < 34$, whereas the bco- and diamond phases
for $f < 44$. It is therefore anticipated that the bco$_2$-, 
bco$_4$- and fcc-phases seen in Fig.\ \ref{latticesums.fig} for
$f < 48$ will be wiped out by the fluid there. However, by arbitrarily
increasing $f$ one can always reach a domain where the fluid will
be beaten by the crystal and hence the bco$_2$ $\to$ bco$_3$ 
$\to$ A15-transitions will be there also after a finite-temperature
calculation. In particular, the A15-phase, on whose stability it has
been speculated for some time, has now been proved to be indeed
the most stable phase at sufficiently high densities 
among the candidates considered.  

The logarithmic-Yukawa potential employed in our study has been
shown to describe well the effective interactions between star polymers
in a good solvent for a wide range of concentrations. However,
at the region of stability of the A15-crystal, the pair potential
description is not expected to be particularly accurate. Many-body
contributions are expected to become important there \cite{arben:triplet}.
Moreover, the Yukawa tail of the potential, describing the interactions
of the outermost Daoud-Cotton blobs, should be absent since the compressed
coronas there are deprived of the outermost blob structure of the 
isolated stars. Thus, in this respect, the logarithmic-Yukawa potential
has to be looked upon rather as a toy model. Nevertheless, 
the physical characteristic driving the transitions discovered above
is the ultraslow divergence of the logarithm which, in the neighborhood
of the average particle distance can be locally expanded as a ramp-like
potential and the finer details of the interaction should become irrelevant.  

\section{Summary and conclusions}
\label{summary.section}

We have shown that ``ultrasoft interactions'' arise naturally
from coarse-graining procedures for a broad range of soft matter
systems.  Besides greatly simplifying the statistical mechanics of
these complex systems -- once the interactions are derived, all the
well known tools of liquid state theory can be applied to calculate
correlations and phase behaviour -- they also lead to new
phenomenology.  Signatures of these ultra soft interactions include
anomalous fluid correlations, reentrant melting as well as the
stabilisation of exotic, open crystal structures. In contrast to
their atomic counterparts, soft matter systems can therefore stabilise
such crystals without the presence of angle-dependent, anisotropic
potentials: radially symmetric, ultrasoft interactions are quite
sufficient. Thus, a new {\it mean field fluid} paradigm can be
established, which goes beyond the usual prototype for classical
fluids, the hard-sphere model. The latter, being always dominated by
packing effects, tends to favour close packed structures. Exotic
crystals with unusual ordering have been observed in hard-sphere-like
suspensions \cite{otewill:prl:92, bartlett:pusey:93} but that case refers 
to {\it binary mixtures} whose phase behaviour is indeed much richer
than that of their one-component counterparts.

\ack
It is with great pleasure that we dedicate this paper 
to Professor Peter Pusey on the occasion of his 60th birthday.
We thank Martin Watzlawek and Primoz Ziherl for helpful discussions.
AAL thanks the Isaac Newton Trust, Cambridge, for financial support.


\section*{References}
\begin{thebibliography}{99}

\bibitem{pusey:leshouches:89} Pusey P N 1991 in {\it Les Houches, 
Session LI, Liquids, Freezing and Glass Transition}, 
edited by J-P Hansen, D Levesque and J Zinn-Justin 
(North-Holland: Amsterdam)

\bibitem{likos:physrep:01} Likos C N 2001 {\it Phys. Rep.} {\bf 348} 267

\bibitem{dijkstra:pre:99} Dijkstra M, van Roij R and Evans R 1999
                          {\it Phys. Rev.} E {\bf 59} 5744

\bibitem{dijkstra:jpcm:99} Dijkstra M, Brader J M and Evans R 1999           
                          {\it J. Phys.: Condens. Matter} {\bf 11} 10079

\bibitem{flory:krigbaum:50} Flory P J and Krigbaum W R 1950 {\it J. Chem.
Phys.} {\bf 18} 1086

\bibitem{deGennes:79} de Gennes P G 1979 {\it Scaling Concepts in Polymer
Physics} (Cornell University Press: Ithaca)

\bibitem{ard:all:02} Louis A A, Bolhuis P G, Finken R, Krakoviack V,
Meijer E J and Hansen J-P 2002 {\it Physica} A {\bf 306} 251

\bibitem{grosberg:82} Grosberg A Y, Khalatur P G and Khokhlov A R 1982
{\it Makromol. Chem. Rapid Commun.} {\bf 3} 709

\bibitem{schaefer:baumgaertner:86} Sch{\"a}fer L and Baumg{\"a}rtner A 1986
{\it J. Phys.} ({\it Paris}) {\bf 47} 1431

\bibitem{dautenhahn:hall:94} Dautenhahn J and Hall C K 1994
{\it Macromolecules} {\bf 27} 5933

\bibitem{krueger:etal:89}
Kr{\"u}ger B, Sch{\"a}fer L and Baumg{\"a}rtner A 1989
{\it J. Phys.} ({\it Paris}) {\bf 50} 3191

\bibitem{ard:peter:00} Louis A A, Bolhuis P G, Hansen J-P and Meijer E J
2000 {\it Phys. Rev. Lett.} {\bf 85} 2522

\bibitem{ard:mft:00} Louis A A, Bolhuis P G and Hansen J-P 2000
{\it Phys. Rev.} E {\bf 62} 7961

\bibitem{bolhuis:jcp:00} Bolhuis P G, Louis A A, Hansen J-P and Meijer E J
2001 {\it J. Chem. Phys.} {\bf 114} 4296

\bibitem{bolhuis:macro:02} Bolhuis P G and Louis A A 2002
{\it Macromolecules} {\bf 35} 1860

\bibitem{bolhuis:pre:01} Bolhuis P G, Louis A A and Hansen J-P 2001
 {\it Phys. Rev.} E {\bf 64} 021801

\bibitem{witten:pincus:86:2} Witten T A and Pincus P A 1986 {\it Macromolecules}
{\bf 19} 2509

\bibitem{jusufi:etal:jpcm:01} Jusufi A, Dzubiella J, Likos C N, von Ferber C
and L{\"o}wen H 2001 {\it J. Phys.: Condens. Matter} {\bf 13} 6177

\bibitem{likos:01} Likos C N, Schmidt M, L{\"o}wen H, Ballauff M, P{\"o}tschke D
and Lindner P 2001 {\it Macromolecules} {\bf 34} 2914

\bibitem{likos:02} Likos C N, Rosenfeldt S, Dingenouts N, Ballauff M, 
Lindner P, Werner N and V{\"o}gtle F 2002 {\it J. Chem. Phys.} in press

\bibitem{Daoud:Cotton:82:1} Daoud M and Cotton J P 1982 {\it J. Phys.} 
({\it Paris})
{\bf 43} 531

\bibitem{likos:etal:prl:98} Likos C N,  L{\"o}wen H, Watzlawek M, Abbas B,
Jucknischke O, Allgaier J and Richter D 1998 {\it Phys. Rev. Lett.} {\bf 80} 4450

\bibitem{jusufi:macromolecules:99} Jusufi A, Watzlawek M and  L{\"o}wen H 1999
{\it Macromolecules} {\bf 32} 4470

\bibitem{arben:prl:02} Jusufi A, Likos C N and L{\"o}wen H 2002 
{\it Phys. Rev. Lett.} {\bf 88} 018301

\bibitem{arben:jcp:02} Jusufi A, Likos C N and L{\"o}wen H 2002
{\it J. Chem. Phys.} in press

\bibitem{evans:79} Evans R 1979 {\it Adv. Phys.} {\bf 28} 143

\bibitem{lang:etal:jpcm:00} Lang A, Likos C N, Watzlawek M and L{\"o}wen H
2000 {\it J. Phys.: Condens. Matter} {\bf 12} 5087

\bibitem{hansen:mcdonald} Hansen J-P and McDonald I R 1986 {\it Theory of
Simple Liquids} 2nd ed (Academic: London)

\bibitem{grewe:klein:jmpa:77} Grewe N and Klein W 1977 {\it J. Math. Phys.}
                              {\bf 64} 1729

\bibitem{grewe:klein:jmpb:77} Grewe N and Klein W 1977 {\it J. Math. Phys.}
                              {\bf 64} 1735

\bibitem{likos:etal:pre:01} Likos C N, Lang A, Watzlawek M and L{\"o}wen H
2001 {\it Phys. Rev.} E {\bf 63} 031206

\bibitem{ard:faraday:01} Louis A A 2001 {\it Phil. Trans. R. Soc. Lond.} A
{\bf 359} 939

\bibitem{archer:evans:pre:01} Archer A J and Evans R 2001 {\it Phys. Rev.} E
{\bf 64} 041501

\bibitem{archer:evans:jpcm:02} Archer A J and Evans R 
2002 {\it J. Phys.: Condens. Matter} {\bf 14} 1131

\bibitem{rogers:young} Rogers F A and Young D A 1984 {\it Phys. Rev.} A
{\bf 30} 999

\bibitem{watzlawek:etal:jpcm:98} Watzlawek M, L{\"o}wen H and Likos C N 
1998 {\it J. Phys.: Condens. Matter} {\bf 10} 8189

\bibitem{stillinger:76} Stillinger F H 1976 {\it J. Chem. Phys.} {\bf 65}
3968

\bibitem{stillinger:weber:78} Stillinger F H and Weber T A  1978
{\it J. Chem. Phys.} {\bf 68} 3837

\bibitem{stillinger:jcp:79} Stillinger F H 1979 {\it J. Chem. Phys.} {\bf 70}
4067

\bibitem{stillinger:prb:79} Stillinger F H 1979 {\it Phys. Rev.} B {\bf 20}
299

\bibitem{stillinger:weber:80} Stillinger F H and Weber T A  1978
{\it Phys. Rev.} B {\bf 22} 3790

\bibitem{stillinger:stillinger:97} Stillinger F H and Stillinger D K 1997
{\it Physica} A {\bf 244} 358

\bibitem{hansen:verlet:69} Hansen J-P and Verlet L 1969 {\it Phys. Rev.}
{\bf 184} 151

\bibitem{hansen:schiff:73} Hansen J-P and Schiff D 1973 {\it Mol. Phys.}
{\bf 25} 1281

\bibitem{likos:psm:98} Likos C N, Watzlawek M and L{\"o}wen H
1998 {\it Phys. Rev.} E {\bf 58} 3135

\bibitem{schmidt:cecam:99} Schmidt M 1999 {\it J. Phys.: Condens. Matter}
{\bf 11} 10163 

\bibitem{fernaud:jcp:00} Fernaud M J, Lomba E and Lee L L 2000
{\it J. Chem. Phys.} {\bf 112} 810

\bibitem{hales:00} Hales T C 2000 {\it Notices Am. Math. Soc.} {\bf 47} 440

\bibitem{hone:83} Hone D, Alexander S, Chaikin P M and Pincus P 1983
{\it J. Chem. Phys.} {\bf 79} 1474

\bibitem{kremer:robbins:grest:86} Kremer K, Robbins M O and Grest G S 1986
{\it Phys. Rev. Lett.} {\bf 57} 2694

\bibitem{robbins:kremer:grest:88} Robbins M O, Kremer K and Grest G S 1988
{\it J. Chem. Phys.} {\bf 88} 3286

\bibitem{sirota:89} Sirota E B, Ou-Yang H D, Sinha S K and Chaikin P M 1989
{\it Phys. Rev. Lett.} {\bf 62} 1524

\bibitem{martin:prl:99} Watzlawek M, Likos C N and L{\"o}wen H 1999
{\it Phys. Rev. Lett.} {\bf 82} 5289

\bibitem{martin:phd:00} Watzlawek M 2000 {\it Phase Behavior of
Star Polymers} (Shaker: Aachen)

\bibitem{witten:pincus:86:1} Witten T A, Pincus P A and Cates M E 1986
{\it Europhys. Lett.} {\bf 2} 137

\bibitem{ashcroft:mermin} Ashcroft N W and Mermin N D 1976
{\it Solid State Physics} (Holt-Saunders: Philadelphia)

\bibitem{ziherl:prl:00} Ziherl P and Kamien R D 2000 {\it Phys. Rev. Lett.}
{\bf 85} 3528

\bibitem{ziherl:jpcb:01} Ziherl P and Kamien R D 2001 {\it J. Phys. Chem.} B
{\bf 105} 10147

\bibitem{kelvin:1887} Thomson W 1887 {\it Philos. Mag.} {\bf 24} 503

\bibitem{weaire:94} Weaire D and Phelan R 1994 {\it Philos. Mag. Lett.}
{\bf 69} 107

\bibitem{rivier:94} Rivier N 1994 {\it Philos. Mag. Lett.} {\bf 69} 297

\bibitem{balag:97} Balagurusamy V S K, Ungar G, Percec V and
Johansson G 1997 {\it J. Am. Chem. Soc.} {\bf 119} 1539

\bibitem{arben:triplet} von Ferber C, Jusufi A, Likos C N, L{\"o}wen H and
Watzlawek M 2000 {\it Eur. Phys. J.} E {\bf 2} 311

\bibitem{otewill:prl:92} Bartlett P, Otewill R H and Pusey P N 1992
{\it Phys. Rev. Lett.} {\bf 68} 3801

\bibitem{bartlett:pusey:93} Bartlett P and Pusey P N 1993 {\it Physica} A
{\bf 194} 415

\end{thebibliography}
\end{document}